\documentclass[aps,prd,preprintnumbers,superscriptaddress,showpacs,floatfix,letterpaper,nofootinbib]{revtex4}
\usepackage{epsfig}
\usepackage{bm}
\usepackage{amssymb}
\usepackage{amsmath}
\usepackage[usenames,dvipsnames]{color}
\usepackage{subfigure}
\usepackage{graphicx}
\usepackage[normalem]{ulem}
\usepackage{cancel}

\begin{document}

\title{Electromagnetic radiation as a probe of the initial state and of
viscous dynamics in relativistic nuclear collisions}
\author{Gojko Vujanovic}
\affiliation{Department of Physics, McGill University, 3600 rue University, Montr\'eal,
Qu\'ebec H3A 2T8, Canada}
\affiliation{Department of Physics, Ohio State University, 191 West Woodruff Avenue,
Columbus, Ohio 43210, USA}
\author{Jean-Fran\c cois Paquet}
\affiliation{Department of Physics, McGill University, 3600 rue University, Montr\'eal,
Qu\'ebec H3A 2T8, Canada}
\affiliation{Department of Physics \& Astronomy, Stony Brook University, Stony Brook, New
York 11794, USA }
\author{Gabriel S. Denicol}
\affiliation{Department of Physics, McGill University, 3600 rue University, Montr\'eal,
Qu\'ebec H3A 2T8, Canada}
\affiliation{Physics Department, Brookhaven National Lab, Building 510A, Upton, New York
11973, USA}
\author{Matthew Luzum}
\affiliation{Departamento de F\'isica de Part\'iculas and IGFAE, Universidade de Santiago
de Compostela, E-15706 Santiago de Compostela, Galicia-Spain}
\affiliation{Instituto de F\'isica - Universidade de S\~ao Paulo, Rua do Mat\~ao Travessa
R, no. 187, 05508-090, Cidade Universit\'aria, S\~ao Paulo, Brasil}
\author{Sangyong Jeon}
\affiliation{Department of Physics, McGill University, 3600 rue University, Montr\'eal,
Qu\'ebec H3A 2T8, Canada}
\author{Charles Gale}
\affiliation{Department of Physics, McGill University, 3600 rue University, Montr\'eal,
Qu\'ebec H3A 2T8, Canada}

\begin{abstract}
The penetrating nature of electromagnetic signals makes them suitable probes
to explore the properties of the strongly-interacting medium created in
relativistic nuclear collisions. We examine the effects of the initial
conditions and shear relaxation time on the spectra and flow coefficients of
electromagnetic probes, using an event-by-event 3+1D viscous hydrodynamic
simulation (\textsc{music}).
\end{abstract}

\maketitle


\section{Introduction}


Ultra-relativistic heavy-ion collisions at the Relativistic Heavy Ion
Collider (RHIC) and the Large Hadron Collider (LHC) are able to reach
temperatures high enough to create and study the quark-gluon plasma (QGP) in
a controlled experimental environment. Experiments performed at these
colliders showed that this new state of matter can be well described by
relativistic hydrodynamics \cite{Heinz:2013th,Gale:2013da}.

Investigating and extracting the properties of the QGP from heavy-ion
collisions is not a simple task since this phase of matter is only created
for a small period of time. Currently, most of our knowledge of the
properties of this hot QCD medium originates from analyzing the momentum
distribution of hadrons. However, the information that can be extracted from
hadrons is limited, since such particles are emitted mainly at the very
late stages of the collision. Indeed, hydrodynamical studies at RHIC
energies have shown that such probes are poorly sensitive to several
properties of the thermalized QCD fluid, such as the initial values of the
shear-stress tensor \cite{Song:2009rh}, the temperature dependence of the
shear viscosity coefficient \cite{Niemi:2011ix}, and the dependence on the
relaxation time \cite{Song:2008si}, among others.

On the other hand, since photons and dileptons are emitted throughout the
entire evolution of the medium, including the QGP sector, they are expected
to carry information about the early stages of the collision and can
potentially be used as probes of the fluid's transport properties. The
PHENIX Collaboration was the first to release measurements of the direct
photon elliptic flow \cite{Adare:2011zr}, soon followed by the ALICE
Collaboration, which released preliminary measurements of the same
observable at LHC energies \cite{Lohner:2012ct}. The main feature observed
by both collaborations is the significant yield of direct photons and their
large azimuthal momentum anisotropy.

Whether hydrodynamical models can be adapted to describe the photon $v_{2}$
is still under investigation. Early hydrodynamical calculations
under-predicted the elliptic flow by factors of $\sim 2-4$~\cite%
{Chatterjee:2013naa,Dion:2011pp,Shen:2013}. Recent calculations using more
complete hadronic photon emission rates and improved modelling of the medium
show a significantly reduced tension between theoretically computed elliptic
flow and the measured $v_{2}$ both at RHIC and the LHC~\cite%
{vanHees:2014ida,Paquet:2015lta}. As far as the dilepton yield is concerned,
theoretical models were able to quantitatively describe the data from the
SPS and RHIC (see \cite{Rapp:1999ej,Rapp:2009yu} for a review). At the time
of this writing, the dielectron elliptic flow had only been measured by the
STAR Collaboration \cite{Adamczyk:2014lpa}, and current theoretical
calculations are consistent with data \cite{Vujanovic:2013jpa}, though large
uncertainties are preventing more definitive conclusions.

In this paper we investigate how electromagnetic probes can complement
hadronic probes in understanding the non-equilibrium dynamics of the hot QCD
in the early stages of the collision. We explicitly demonstrate that thermal
photons and dileptons emitted by the QGP are sensitive to the shear
relaxation time --- a transport coefficient that has a negligible effect on
hadronic observables. We further study the effects that a non-equilibrium
initial condition can have on EM probes. Thermal photons and dileptons are
found to be affected by the non-equilibrium aspects of the initial profile. 

\section{Modeling the evolution of the medium created at RHIC}

\label{Sec:modeling_medium} 

\subsection{Initial condition}

The hydrodynamical evolution starts at time $\tau _{0}=0.4$~fm with an
initial energy density profile given by the Glauber model. In our
implementation of this model, the initial energy density profile is assumed
to be factorized into a longitudinal part and a transverse part, as
originally proposed in Ref. \cite{Hirano:2002ds}: 
\begin{equation*}
\varepsilon \left( \tau _{0},x,y,\eta \right) =\exp \left[ -\frac{\left(
\left\vert \eta \right\vert -\eta _{\mathrm{flat}}/2\right) ^{2}}{2\eta
_{\sigma }^{2}}\theta \left( \left\vert \eta \right\vert -\eta _{\mathrm{flat%
}}/2\right) \right] \varepsilon _{T}\left( x,y\right) .
\end{equation*}

In the longitudinal $\eta $-direction, we take a profile that is
approximately boost invariant near mid-rapidity and falls like a Gaussian at
large rapidities. We set $\eta _{\mathrm{flat}}=5.9$ and $\eta _{\sigma
}=0.4 $. The energy density in the transverse direction is given by the
Monte Carlo Glauber (MC Glauber) model 
\begin{equation*}
\varepsilon _{T}\left( x,y\right) =W\left[ \alpha n_{\mathrm{BC}}\left(
x,y\right) +\left( 1-\alpha \right) n_{\mathrm{WN}}\left( x,y\right) \right]
,
\end{equation*}%
where we defined 
\begin{equation*}
n_{\mathrm{BC/WN}}\left( x,y\right) =\frac{1}{2\pi \sigma ^{2}}%
\sum_{i=1}^{N_{\mathrm{bin/part}}}\exp \left[ -\frac{\left( x-x_{i}\right)
^{2}+\left( y-y_{i}\right) ^{2}}{2\sigma ^{2}}\right] ,
\end{equation*}%
with $W$ being an overall normalization factor, and $\alpha $ a parameter
dictating the proportion in which wounded nucleons and binary collisions
contribute to the energy density profile in the transverse plane.
Furthermore, $N_{\mathrm{part}}$ and $N_{\mathrm{bin}}$ are the number of
participants and binary collision of a given event and ($x_{i}$,$y_{i}$) are
the coordinates of the corresponding participant or binary collision on the
transverse plane. In this work, we set $W=6.16$ GeV/fm and $\alpha =0.25$
for all simulations. The number and coordinates of participants and binary
collisions are calculated taking a nucleon-nucleon inelastic cross section
of $\sigma _{NN}=42.1$ mb. The fluctuation scale $\sigma $, which specifies
the length scale of energy density fluctuations, is taken to be $\sigma =0.4$
fm. The above parameters are based on Ref. \cite{Schenke:2011bn}, but were adjusted to fit 
the charged pion transverse momentum spectrum and charged hadron elliptic flow 
at $\sqrt{s_{NN}}=200$ GeV at RHIC in the 20\%-40\% centrality class.

We select the centrality class by sampling events in a certain range of
impact parameters. For the $20$--$40$\% centrality class considered in this
paper, we sampled events with impact parameters, $b_{\mathrm{imp}}$, ranging
from $b_{\mathrm{imp}}=6.7$--$9.48$ fm. A total of two hundred events were
sampled.

Finally, we provide initial conditions for the 4-velocity, $u^{\mu }$, and
shear-stress tensor, $\pi ^{\mu \nu }$. The initial flow profile is set to
be zero, i.e., $u_{0}^{\mu }=(1,0,0,0)$ in $(\tau ,\eta )$ coordinates,
while the initial shear-stress tensor is always assumed to be proportional
to its corresponding Navier-Stokes value, 
\begin{equation*}
\pi ^{\mu \nu }(\tau _{0})=c\times \mathrm{diag}\left( 0,\frac{2\eta }{3\tau
_{0}},\frac{2\eta }{3\tau _{0}},-\frac{4\eta }{3\tau _{0}}\right) ,
\end{equation*}%
where the parameter $c$ is a constant and will be varied between $0$ and $1$
in this work.

\subsection{Relativistic dissipative fluid dynamics}

The time evolution of the system is described using relativistic dissipative
fluid dynamics. The evolution of the energy-momentum tensor $T^{\mu \nu }$
is first constrained by the conservation law: 
\begin{equation}
\partial _{\mu }T^{\mu \nu }=0,  \label{1}
\end{equation}%
where $T^{\mu \nu }=\varepsilon u^{\mu }u^{\nu }-\Delta ^{\mu \nu }P+\pi
^{\mu \nu }$ with $P$ being the thermodynamic pressure and $\Delta ^{\mu \nu
}=g^{\mu \nu }-u^{\mu }u^{\nu }$ the projection operator onto the 3-space
orthogonal to velocity. We employ Landau's definition of the velocity field 
\cite{landau} and assume that bulk viscous pressure and baryon number
4-current are identically zero in all space-time points. The equation of
state, which dictates how the thermodynamic pressure changes as a function
of energy density, is taken from Ref.~\cite{Huovinen:2009yb} and corresponds
to a parametrization of a lattice QCD calculation, at high temperatures,
smoothly connected to a parametrization of the hadron resonance gas at lower
temperatures. At temperatures below $T_{\mathrm{ch}}=0.16$ GeV, this
equation of state follows a partial chemical equilibrium prescription, which
assumes that ratios of particle multiplicity remain fixed for all $T<T_{%
\mathrm{ch}}$ \cite{Bebie:1991ij,Hirano:2002ds}.

The evolution equation for the shear-stress tensor is provided by
Israel-Stewart theory \cite{Israel:1979wp,Israel1976310}, 
\begin{equation}
\tau _{\pi }\Delta _{\alpha \beta }^{\mu \nu }u^{\lambda }\partial _{\lambda
}\pi ^{\alpha \beta }+\pi ^{\mu \nu }=2\eta \sigma ^{\mu \nu }+\frac{4}{3}%
\tau _{\pi }\pi ^{\mu \nu }\partial _{\lambda }u^{\lambda }\text{,}
\label{eq:pi_munu}
\end{equation}%
where $\sigma ^{\mu \nu }=\Delta _{\alpha \beta }^{\mu \nu }\partial
^{\alpha }u^{\beta }$ is the shear tensor and $\Delta _{\alpha \beta }^{\mu
\nu }=\left( \Delta _{\alpha }^{\mu }\Delta _{\beta }^{\nu }+\Delta _{\beta
}^{\mu }\Delta _{\alpha }^{\nu }\right) /2-\left( \Delta _{\alpha \beta
}\Delta ^{\mu \nu }\right) /3$ is the double, symmetric, traceless
projection operator. Above, we introduced two transport coefficients, the
shear viscosity coefficient, $\eta $, and the shear relaxation time, $\tau
_{\pi }$. In principle, additional nonlinear terms exist in the
Israel-Stewart theory \cite{Denicol:2010xn,Denicol:2012cn}, but for the sake
of simplicity, they are not included in this work.

The transport coefficients are functions of the temperature (and the baryon
chemical potential, if the baryon number density is nonzero) that, in
principle, should be computed from the underlying microscopic theory.
However, reliable calculations of the aforementioned transport coefficients
in the strongly coupled regime are not yet possible. In this work, we assume
the existence of an effective shear viscosity coefficient that is
proportional to the entropy density, 
\begin{equation*}
\frac{\eta }{s}=0.08\text{.}
\end{equation*}
Meanwhile, the relaxation time is assumed to be of the form, 
\begin{equation}
\tau _{\pi }=b_{\pi }\frac{\eta }{\varepsilon +P},  \label{eq:taupi}
\end{equation}
with $b_{\pi }$ being varied from 5 to 20. We note that, in order to
preserve causality, the coefficient $b_{\pi}$ is constrained to be $b_{\pi}
\geq 4/[3\left( 1-c_{s}^{2}\right) ]$, where $c_{s}$ is the velocity of
sound \cite{Pu:2009fj}.

The fluid-dynamical equations are solved numerically using the \textsc{music
2.0} simulation code, an updated version of the simulation code presented in
Ref. \cite{Schenke:2010nt,Schenke:2010rr,Schenke:2011bn}. This simulation
code has recently been tested against semi-analytic solutions of
Israel-Stewart theory and was shown to provide accurate solutions of such
type of equations \cite{Marrochio:2013wla}. The simulations performed in
this paper used a time step of $\Delta \tau =0.03$ fm and a grid spacing of $%
\Delta x=\Delta y=1/6$ fm and $\Delta \eta =1/5$. Such values are small
enough to ensure that we achieved the continuum limit for the particular
observables explored in this study.


\subsection{Particle production}

Hadrons are produced using the traditional Cooper-Frye prescription \cite%
{Cooper:1974mv}, with a constant temperature freeze-out hypersurface,
defined by $T_{FO}=145$ MeV, and including all 2-- and 3--particle decays of
hadronic resonances \cite{Sollfrank:1991xm} up to 1.3 GeV.

In the Cooper-Frye formalism, one first needs to specify the local momentum
distribution of hadrons. For an ideal hadron resonance gas, these would
correspond to Fermi-Dirac or Bose-Einstein distributions, with the
appropriate mass and degeneracy factor. For dissipative systems, this is no
longer the case and the distribution function should be generalized to also
include non-equilibrium corrections. Here, we use \cite{Teaney:2003kp} 
\begin{equation}
f_{\mathbf{k}}^{i}=f_{0\mathbf{k}}^{i}+\delta f_{\mathbf{k}%
}^{i};\;\;\;\delta f_{\mathbf{k}}^{i}=f_{0\mathbf{k}}^{i}\left( 1+af_{0%
\mathbf{k}}^{i}\right) \frac{\pi ^{\mu \nu }}{2\left( \varepsilon +P\right)
T^{2}}k_{\mu }^{i}k_{\nu }^{i},  \label{eq:df}
\end{equation}%
where the index $i$ specifies the hadron species, $k_{\mu }^{i}$ is that
hadron's $4$--momentum, and $a=1\left( -1\right) $ for bosons (fermions).


\section{Production rates of Electromagnetic Probes}

\label{Sec:EM_probes} 

The production rate of electromagnetic radiation from a QCD plasma is known
only in very specific limits of the QCD phase diagram. At low temperatures,
the emission rate has been described using effective Lagrangians with
hadronic degrees of freedom \cite{Turbide:2003si,Rapp:1999ej,Eletsky:2001bb,Lee:1998um}. For weakly-coupled
and high temperature plasma, the rate has been computed perturbatively at
next-to-leading order for photons~\cite{Ghiglieri:2013gia} and dileptons \cite{Laine:2013vma,Ghisoiu:2014mha,Ghiglieri:2014kma}. 
Electromagnetic emission rates that take into account deviations from local thermodynamic equilibrium 
--- essential when the emitting medium is a viscous fluid ---
have been published recently \cite{Schenke:2006yp,Dusling:2008xj,Dion:2011pp,Shen:2013,Vujanovic:2013jpa},
although they have not yet been extended to include next-to-leading order
processes.

Since EM rates are known only for the low and high temperature limits of the
QCD medium, the approach taken in this paper is to use each rate in their
own temperature limits. In the crossover region, here taken to be in the
temperature range $T=184$--$220$ MeV, we use rates that are a linear
interpolation of hadronic and QGP rates.

For the strong coupling constant, we used a constant value of $g_s=2$,
corresponding to $\alpha_s\approx0.32$. The rates used for photons and
dileptons are described in more details in the following subsections.


\subsection{Dileptons}

\label{dileptons_explanation} 

The dilepton production rate can be written as 
\begin{equation}
\frac{d^{4}R^{\ell ^{+}\ell ^{-}}}{d^{4}q}=-\frac{L(M)}{M^{2}}\frac{%
\alpha_{EM}^{2}}{\pi ^{3}}\frac{\mathrm{Im}\Pi _{EM}^{R}(M,|\mathbf{q}|;T)}{%
e^{q^{0}/T}-1} \; ; \; \; L(M)=\left( 1+\frac{2m_{\ell}^{2}}{M^{2}}\right) 
\sqrt{1-\frac{4m_{\ell }^{2}}{M^{2}}} \text{ ,}  \label{eq:dilep_rate}
\end{equation}%
\ where $M^2=q_{\mu }q^{\mu }$, $q^{0}=\sqrt{M^{2}+|\mathbf{q}|^{2}}$, and $%
\mathrm{Im}\Pi _{EM}^{R}$ is the imaginary part of the retarded virtual
photon self-energy.

At low temperatures, we use the Vector meson Dominance Model (VDM), first
proposed by Sakurai \cite{Gounaris:1968mw}, to relate the real and virtual
photon self-energy to hadronic degrees of freedom. According to the VDM, the
imaginary part of the retarded photon self-energy $\mathrm{Im}\Pi _{EM}^{R}$
is related to the imaginary part of the retarded vector meson propagator $%
\mathrm{Im}D_{V}^{R}$ via 
\begin{equation}
\mathrm{Im}\Pi _{EM}^{R}=\sum_{V=\rho ,\omega ,\phi }\left( \frac{m_{V}^{2}}{%
g_{V}}\right) ^{2}\mathrm{Im}D_{V}^{R}\text{ ,}
\end{equation}%
where $V=\rho ,\omega ,\phi $ denote the corresponding vector mesons, and $%
g_{V}$ is the coupling constant between a vector meson $V$ and a photon. In
the Schwinger-Dyson formalism the retarded propagator can be related to the
vector meson self energy \cite{Roberts:1994dr}. The finite temperature piece
of the self-energy has been computed through the forward scattering
amplitude method first devised by Eletsky~\textit{et~al.}~\cite%
{Eletsky:2001bb}, which includes two contributions: i) resonance scatterings
through experimentally observed particles and ii) non-perturbative Regge
physics. The vacuum piece of the self-energy is computed using chiral
effective Lagrangians. Recently, this method has been extended to include
viscous corrections \cite{Vujanovic:2013jpa} assuming that viscosity
modifies the thermal particle distribution according to Eq. (\ref{eq:df}).
The full self-energy can be expressed as $\Pi =\Pi _{0}+\delta \Pi $, with $%
\delta \Pi $ being responsible for the viscous corrections to the dilepton
rate.

Another approach is to use chiral effective Lagrangians \cite{Rapp:1999ej}
to study the in-medium properties of vector mesons. However, the approach in
Ref.~\cite{Rapp:1999ej} has not yet been generalized to a viscous
description of the medium, and hence will not be used in this study.

The QGP dilepton emission rate used in this work is calculated from kinetic
theory using the Born approximation, since viscous corrections to this rates
are known. The rate is given by: 
\begin{eqnarray}
\frac{d^{4}R^{\ell ^{+}\ell ^{-}}}{d^{4}q} &=&\int \frac{d^{3}p_{1}d^{3}p_{2}%
}{(2\pi )^{6}E_{\mathbf{p}_{1}}E_{\mathbf{p}_{2}}}f_{\mathbf{p}_{1}}^{(1)}f_{%
\mathbf{p}_{2}}^{(2)}\frac{q^{2}}{2}\sigma \delta ^{4}(q-p_{1}-p_{2})\text{ }%
,  \notag \\
\sigma &=&\frac{16\pi \alpha _{\mathrm{EM}}^{2}\left( \sum_{q^{\prime
}}e_{q^{\prime }}^{2}\right) N_{c}}{3q^{2}}\text{ },
\end{eqnarray}
where $\sigma $ is the cross section for $q+\bar{q}\rightarrow \ell
^{-}+\ell ^{+}$. The QCD number of colors is denoted $N_{c}$, and the sum
over $q^{\prime }$ spans over the quark flavors, which we limit to the three
lightest: $q^{\prime }=u,d,s$. Naturally, the single particle momentum
distribution functions, $f_{\mathbf{p}}^{i}$, includes viscous corrections,
which we assume to be of the same form as the one shown in Eq.~(\ref{eq:df}%
). The QGP rate can thus be expressed as: 
\begin{eqnarray}
\frac{d^{4}R^{\ell ^{+}\ell ^{-}}}{d^{4}q} &=&\frac{d^{4}R_{0}^{\ell
^{+}\ell ^{-}}}{d^{4}q}+\frac{d^{4}\delta R^{\ell ^{+}\ell ^{-}}}{d^{4}q} 
\notag \\
\frac{d^{4}\delta R^{\ell ^{+}\ell ^{-}}}{d^{4}q} &=&\frac{q_{\mu }q_{\nu
}\pi ^{\mu \nu }}{2T^{2}(\varepsilon +P)}b_{2}(q^{0},|\mathbf{q}|,T)
\end{eqnarray}%
where the expression for $b_{2}(q^{0},|\mathbf{q}|,T)$ can be found in \cite%
{Dusling:2008xj,Vujanovic:2013jpa}.


\subsection{Photons}

\label{photons_explanation} 

At low temperatures, we assume that the photon production can be computed by
effectively describing the medium as a gas of light mesons \cite%
{Turbide:2006zz,Turbide:2003si}, via a massive Yang-Mills Lagrangian coupled
with the Vector Dominance Model\footnote{
The contribution from baryons and $\pi \pi $ bremsstrahlung, which are known
to be significant (see e.g. Ref.~\cite{Paquet:2015lta}), are not included
since viscous $\delta f_{\mathbf{p}}$ corrections are not yet known for these rates.%
}. In this case, thermal photon production can be computed within a kinetic
description: 
\begin{eqnarray}
k\frac{d^{3}R^{\gamma }}{d^{3}\mathbf{k}} &=&\int \frac{d^{3}p_{1}}{2E_{%
\mathbf{p}_{1}}(2\pi )^{3}}\frac{d^{3}p_{2}}{2E_{\mathbf{p}_{2}}(2\pi )^{3}}%
\frac{d^{3}p_{3}}{2E_{\mathbf{p}_{3}}(2\pi )^{3}}\frac{1}{2(2\pi )^{3}} 
\notag \\
&&\times (2\pi )^{4}\delta ^{(4)}(P_{1}+P_{2}-P_{3}-K)\left\vert \mathcal{M}%
\right\vert ^{2}f_{\mathbf{p}_{1}}^{(1)}f_{\mathbf{p}_{2}}^{(2)}\left[ 1\pm
f_{\mathbf{p}_{3}}^{(3)}\right] \text{ }.  \label{eq:HG_Rate}
\end{eqnarray}%
This formula corresponds to a linearized Boltzmann equation for $%
2\rightarrow 2$ process, with $\mathcal{M}$ being the zero-temperature
matrix element corresponding to the photon emission process. Contributions
due to photon absorption by the medium are neglected. Some particle decays ($%
1\rightarrow 3$ processes), are included as well in our calculation, with
Eq. (\ref{eq:HG_Rate}) modified accordingly for such process. The deviation
from local equilibrium is taken into account through the presence of $\delta
f_{\mathbf{p}}^{i}$, as in Eq.~(\ref{eq:df}), in the hadron momentum
distribution functions $f_{\mathbf{p}}^{i}$.

At high temperatures, we use the photon emission rate of a
weakly-interacting QGP. We use the full leading order rate as computed in 
\cite{Arnold:2001ba}, which include photon production through Compton
scattering, quark-antiquark annihilation and soft bremsstrahlung. For the
first two processes, we further include the correction due to the
anisotropic momentum distribution associated with the use of viscous
hydrodynamics \cite{Shen:2013}.

For both the QGP and hadron gas, the emission rates can be written in the
form 
\begin{equation}
k\frac{d^{3}R^{\gamma }}{d^{3}\mathbf{k}} = k\frac{d^{3}R^{\gamma }_{0}}{%
d^{3}\mathbf{k}} + k\frac{d^{3}\delta R^{\gamma }}{d^{3}\mathbf{k}}
\end{equation}
which allows for a straightforward separation of the ideal and viscous
contributions to the rate, with the viscous correction $\delta R \propto 
\frac{\pi^{\mu\nu} K_\mu K_\nu}{(\varepsilon+P)}$ given in Ref. \cite%
{Shen:2013}.


\section{Results and Discussion}

\label{Sec:Results} 


\subsection{Effect of the shear relaxation time on EM probes}

\label{discussionTau}

In this section we investigate the effect of the shear relaxation time $\tau
_{\pi }$ on EM probes. The shear relaxation time dictates the time it takes
for the shear stress tensor $\pi ^{\mu \nu }$ to relax towards the
Navier-Stokes (NS) value, $\pi _{NS}^{\mu \nu }=2\eta \sigma ^{\mu \nu }$,
following Eq. (\ref{eq:pi_munu}). The effects of $\tau _{\pi }$ are studied
using the parametrization of the relaxation time given by Eq.~(\ref{eq:taupi}%
). Different relaxation times are modeled through the parameter $b_{\pi }$,
which here is chosen to have three possible values $b_{\pi }=5,10,20$. The
value $b_{\pi }=5$ has been obtained in kinetic theory \cite{Denicol:2012cn}%
. The initialization of $\pi ^{\mu \nu }$ is taken to be $\pi ^{\mu \nu
}(\tau _{0})=0$. Since hadrons are produced predominantly during the later
phase of the medium evolution, the shear relaxation time is only expected to
affect them if $\tau _{\pi }$ is of the order of the medium lifetime. On the
other hand electromagnetic probes are produced throughout the evolution and
may display a larger sensitivity to the value of the relaxation time. 
\begin{figure}[!h]
\begin{center}
\includegraphics[width=0.52\textwidth]{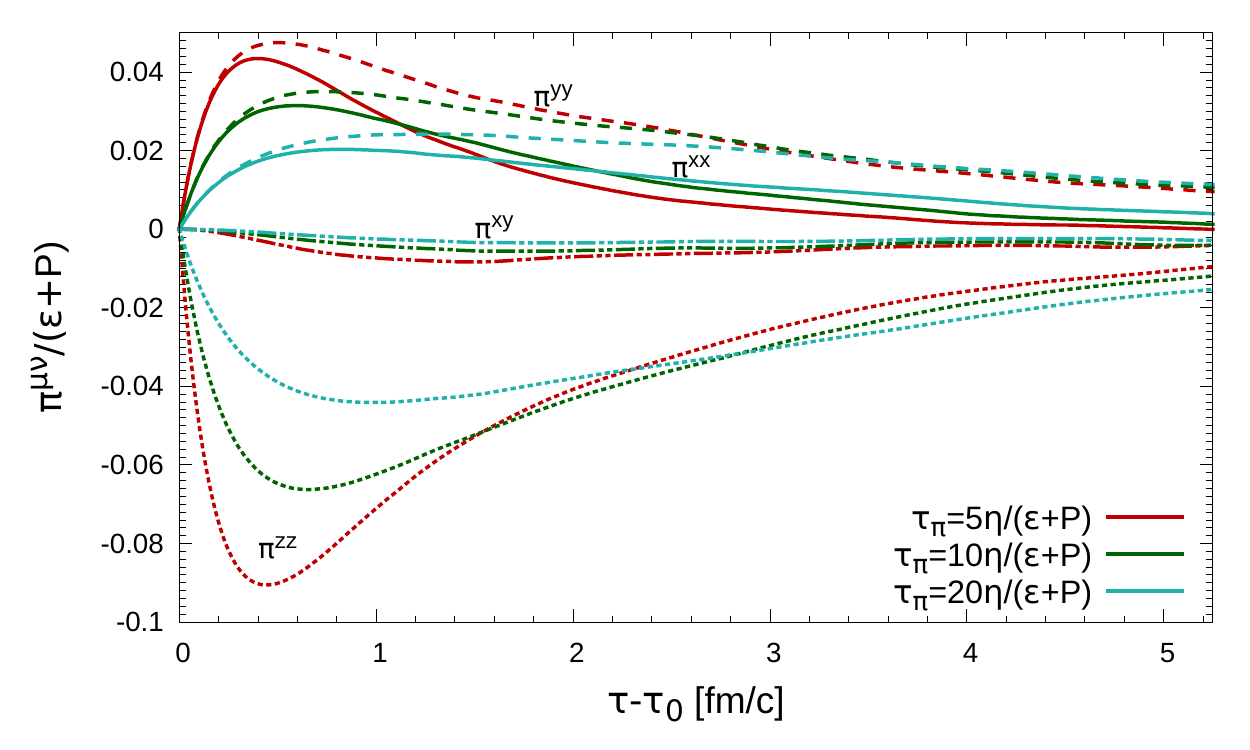}
\end{center}
\caption{(Color online) Event averaged shear-stress tensor for $b_{\protect%
\pi }=5,10,20$ in the local rest frame of the fluid cell located at
x=y=2.625 fm, z=0 fm. Results with $b_{\protect\pi }=5$ are in red, $b_{%
\protect\pi }=10$ in dark green and $b_{\protect\pi }=20$ in light green.
Different types of lines are used for various components of $\protect\pi ^{%
\protect\mu \protect\nu }$. The full line is reserved for $\protect\pi ^{xx}$%
, dashed line for $\protect\pi ^{yy}$, dotted line for $\protect\pi ^{zz}$
and dash-dotted line is used for $\protect\pi ^{xy}$.}
\label{fig:pi_munu_tau_pi}
\end{figure}
We begin by looking at the dimensionless ratio $\bar{\pi}^{\mu \nu }=\pi
^{\mu \nu }(\tau )/(\varepsilon +P)$ in the local rest frame for the three
values of $\tau _{\pi }$. Averaged over 200 hydrodynamical events for each value of $\tau_\pi$, this
ratio is shown in Fig.~\ref{fig:pi_munu_tau_pi} for a given transverse
position (see caption of Fig.~\ref{fig:pi_munu_tau_pi}). Due to the large longitudinal gradients
present at early times, $\bar{\pi}^{\mu \nu }$ rapidly increases in the
first $0.5$ fm of evolution. Given the initial conditions $\bar{\pi}(\tau
_{0})=0$, increasing $\tau _{\pi }$ postpones the build-up of $\bar{\pi}%
^{\mu \nu }(\tau )$. On the other hand, increasing the relaxation time also
reduces the decay rate of $\bar{\pi}^{\mu \nu }(\tau )$ at late times,
leading to relatively larger values of $\bar{\pi}^{\mu \nu }(\tau )$ at long
times.

\begin{figure}[!h]
\par
\begin{center}
\subfigure[]{\includegraphics[width=0.497%
\textwidth]{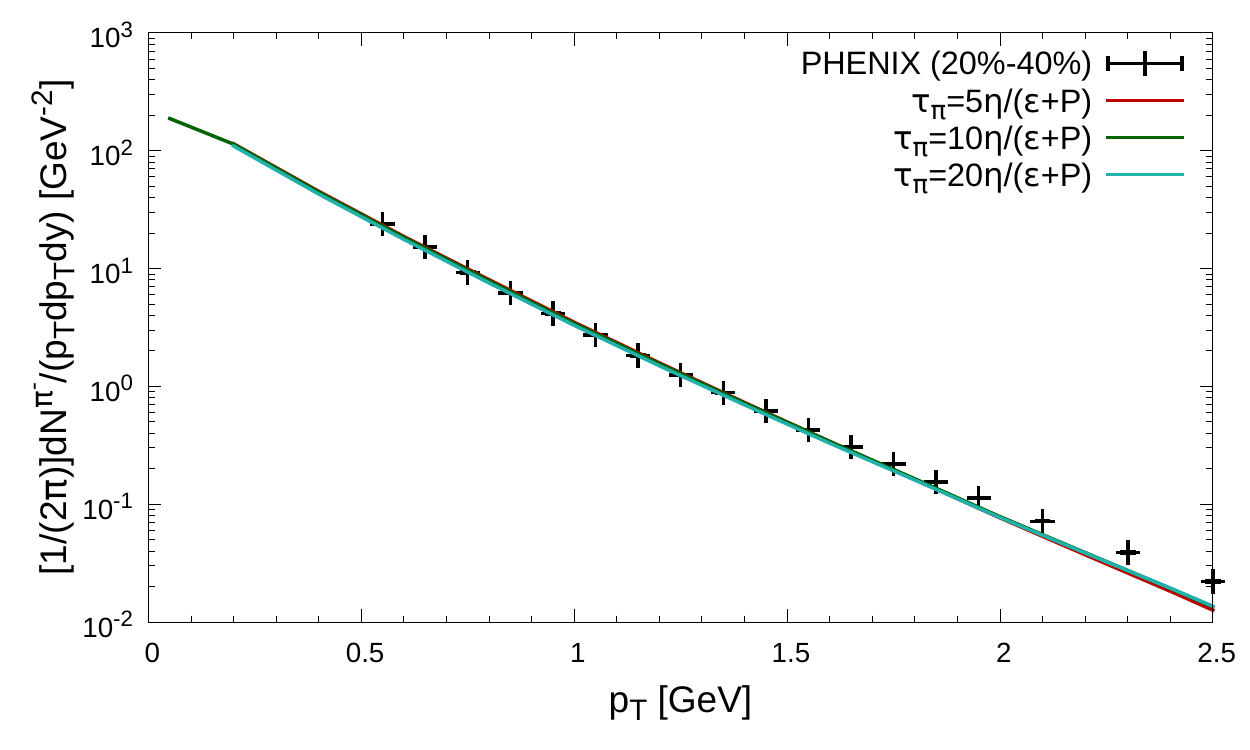}} %
\subfigure[]{\includegraphics[width=0.497%
\textwidth]{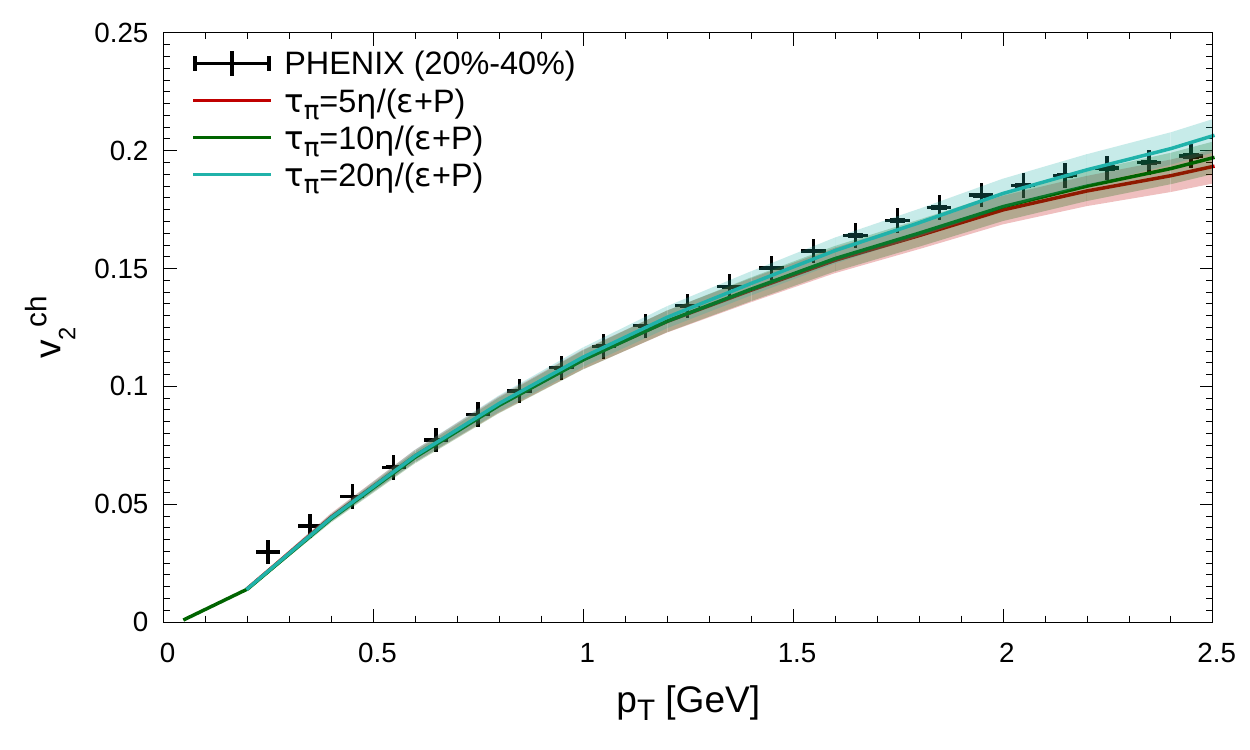}}
\end{center}
\caption{(Color online) Pion transverse momentum spectra (a) and charged
hadron differential elliptic flow (b) as a function of transverse momentum,
for different values of shear relaxation time. Here, and in all subsequent
figures of this paper, the colored bands represent the statistical
uncertainty associated with 200 hydrodynamical events.}
\label{fig:hadrons}
\end{figure}
In Fig.~\ref{fig:hadrons}, we show the pion transverse momentum spectra (a)
and the elliptic flow of charged hadrons (b) for our three choices of
relaxation time\footnote{%
The $v_2$ of charged hadrons is always computed using the root mean square
value of the $v_2$ computed in each of the 200 hydrodynamical simulations.}.
We can see that changes in the relaxation time have little effect on these
observables. This is consistent with the small differences seen in $%
\pi^{\mu\nu}(\tau_{\mathrm{fo}})$ at late times. 
\begin{figure}[!h]
\begin{center}
\subfigure[]{\includegraphics[width=0.497%
\textwidth]{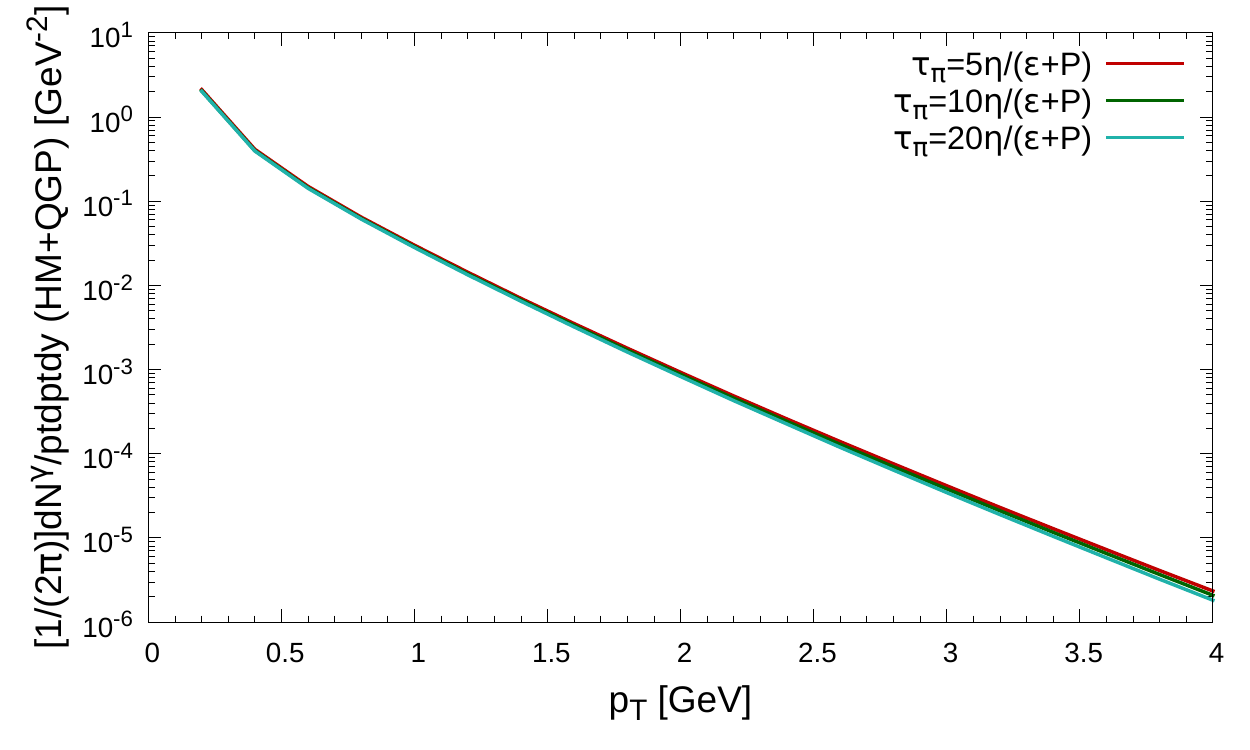}} %
\subfigure[]{\includegraphics[width=0.497%
\textwidth]{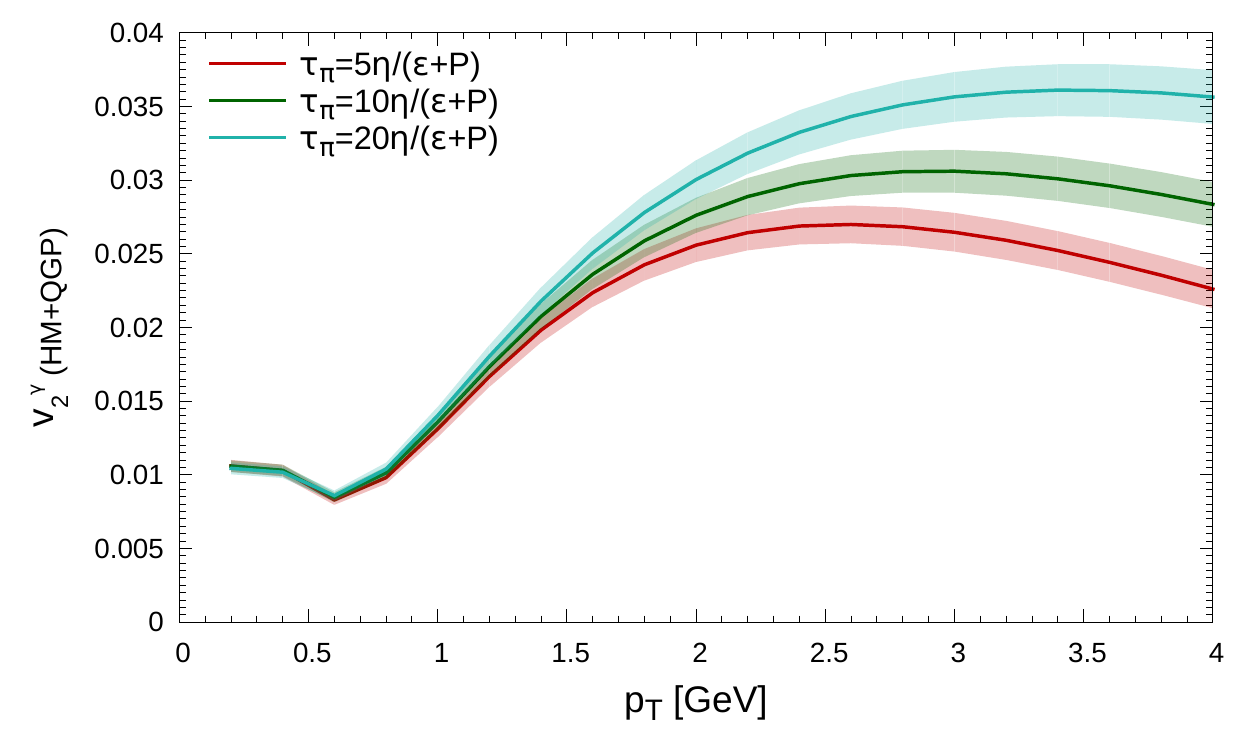}}
\end{center}
\caption{(Color online) Transverse momentum spectra (a) and differential
elliptic flow (b) of thermal photons as a function of transverse momentum,
for different values of shear relaxation time.}
\label{fig:thermal_photon}
\end{figure}
In Fig.~\ref{fig:thermal_photon}, we show the spectra and elliptic flow for
thermal photons. We computed the photon elliptic flow using the scalar
product method: 
\begin{equation}
v^\gamma_n=\frac{\left\langle v^{h}_{n} v^\gamma_{n} \cos \left[%
n\left(\Psi^\gamma_{n}-\Psi^h_{n} \right)\right] \right\rangle_{\mathrm{ev}}%
}{\sqrt{\left\langle (v^h_{n})^2 \right\rangle_{\mathrm{ev}}}}
\label{eq:vnSP}
\end{equation}
where $\langle \ldots \rangle_{\mathrm{ev}}$ is an average over events. The $%
v^s_n$ and $\Psi^s_n$ in single event are given by 
\begin{equation}
v^s_n e^{i n \Psi^s_n} = \frac{\int d p_T dy d\phi p_T \left[ p^0 \frac{d^3
N^s}{d^3 p} \right] e^{i n\phi}}{\int d p_T dy d\phi p_T \left[ p^0 \frac{%
d^3 N^s}{d^3 p} \right]}
\end{equation}
where $p^0 d^3 N^s/d^3 p$ is the single-particle distribution of particle
species $s$. The hadronic $v^h_n$ and $\Psi^h_{n}$ used in Eq. (\ref{eq:vnSP}%
) are integrated over $-0.35<\eta<0.35$ and $0.035<p_T<3$~GeV to simulate
the large bin used experimentally. The photon $v^{\gamma}_n$ and $%
\Psi^{\gamma}_{n}$ are evaluated at mid-rapidity, for given values of $p_T$.
The dilepton anisotropies are computed using the same approach, with the
more general single-particle distribution $d^4 N^s/d^4 p$.

The photon yield is slightly reduced when increasing $b_\pi$ from $5$ to $20$%
. We have verified that the source of this change originates from a
reduction of the viscous correction ($\delta$R) to the photon production
rate, which is proportional to $\bar{\pi}^{\mu\nu}(\tau)$ and thus decreases
as $\tau_\pi$ increases, as seen in Fig.~\ref{fig:pi_munu_tau_pi}. 
\begin{figure}[!th]
\begin{center}
\subfigure[]{\includegraphics[width=0.497%
\textwidth]{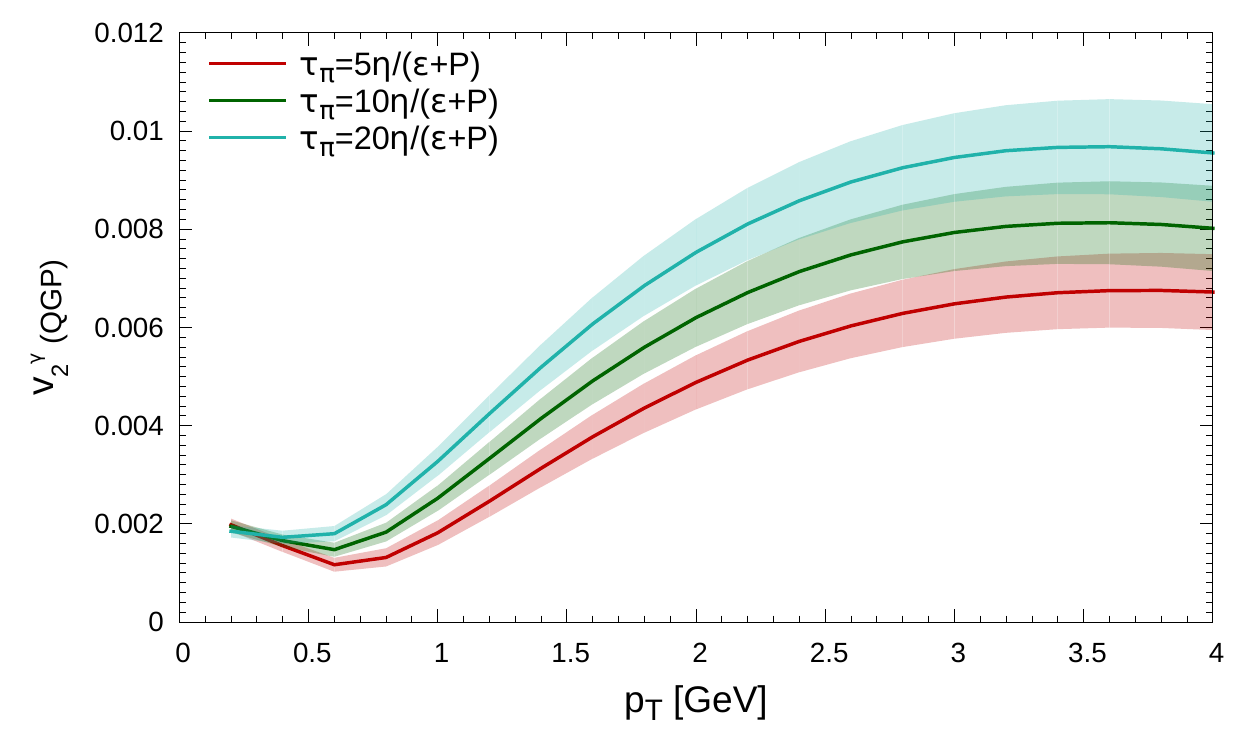}} %
\subfigure[]{\includegraphics[width=0.497%
\textwidth]{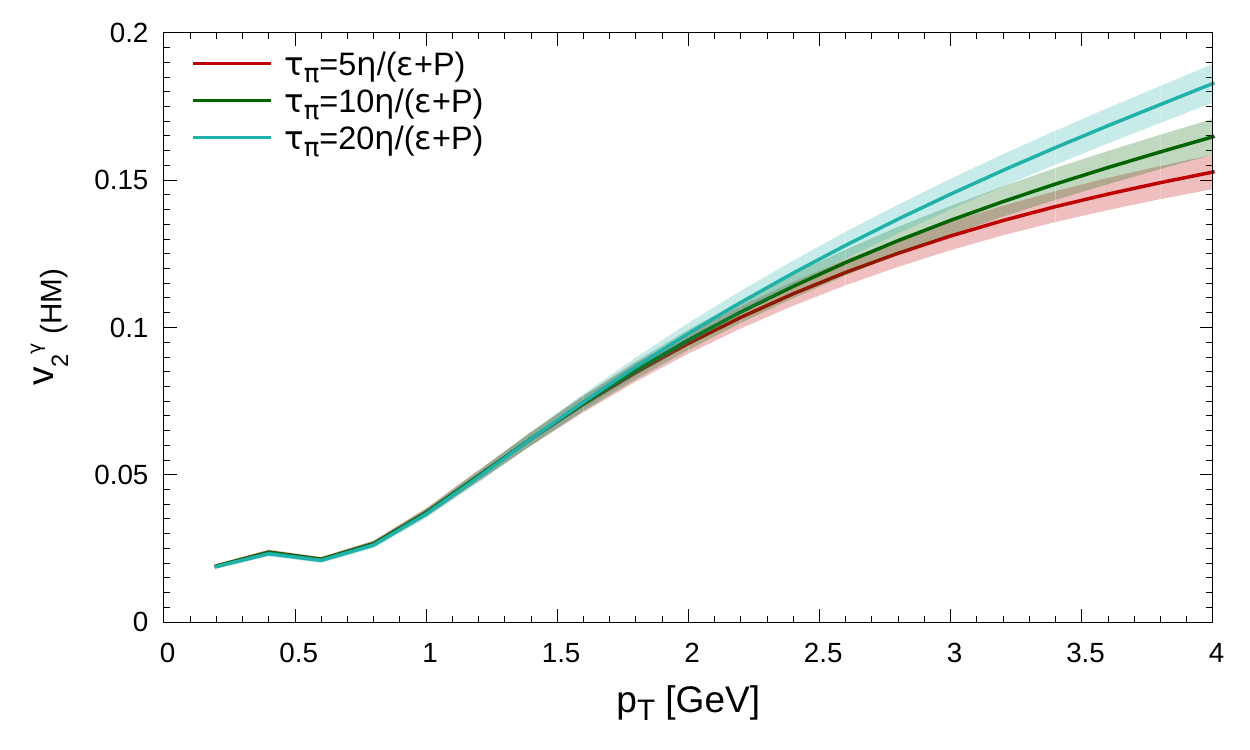}}
\end{center}
\caption{(Color online) Differential elliptic flow of thermal photons
emitted by the QGP (a) and emitted by the hadronic medium (HM) (b) as a
function of transverse momentum, for different values of shear relaxation
time.}
\label{fig:photon_hg_qgp_v2}
\end{figure}

On the other hand the elliptic flow of photons is increased by more than
50\% for $p_{T}>$ 3 GeV. The contribution from the individual sources is
isolated in Fig.~\ref{fig:photon_hg_qgp_v2}, where the elliptic flow of
thermal photons produced in the QGP phase is in panel (a) while the one
originating from photons produced in the hadronic medium (HM) phase is in panel (b). The
elliptic flow of thermal photons emitted in the hadronic stage of the
evolution is not significantly affected by the relaxation time, while the
elliptic flow originating from the QGP thermal photons displays a large
dependence on $\tau _{\pi }$. Such strong dependence on the relaxation time
remains in the total thermal photon $v_{2}$ for $p_{T}\gtrsim 1.5$ GeV,
since the total $v_{2}$ is a delicate balance --- a yield
weighted average --- of the individual sources.

Therefore, the overall dependence of thermal photons on the relaxation time
is not universal, since it depends on a nontrivial balancing between thermal
probes emitted at early times (which corresponds roughly to QGP emissions)
and thermal probes emitted at later times from the HM. In this sense, it is
possible that thermal dileptons' invariant mass distribution, integrated
over $p_{T}$, is better suited to see this behavior. At small invariant
masses $M\lesssim 1.1$ GeV, the HM thermal emission dominates. As the
invariant mass increases, the relative contribution of QGP dileptons
gradually becomes more pronounced, finally dominating at intermediate
invariant masses $M\gtrsim 1.1$ GeV. However, the dominance of HM dileptons
at low invariant masses is not $p_{T}$ independent. Hence before describing
the invariant mass $M$ distributions of thermal dileptons, let us first
consider the $p_{T}$ distribution at fixed $M$.

\begin{figure}[!h]
\centering
\subfigure[]{\includegraphics[width=0.497%
\textwidth]{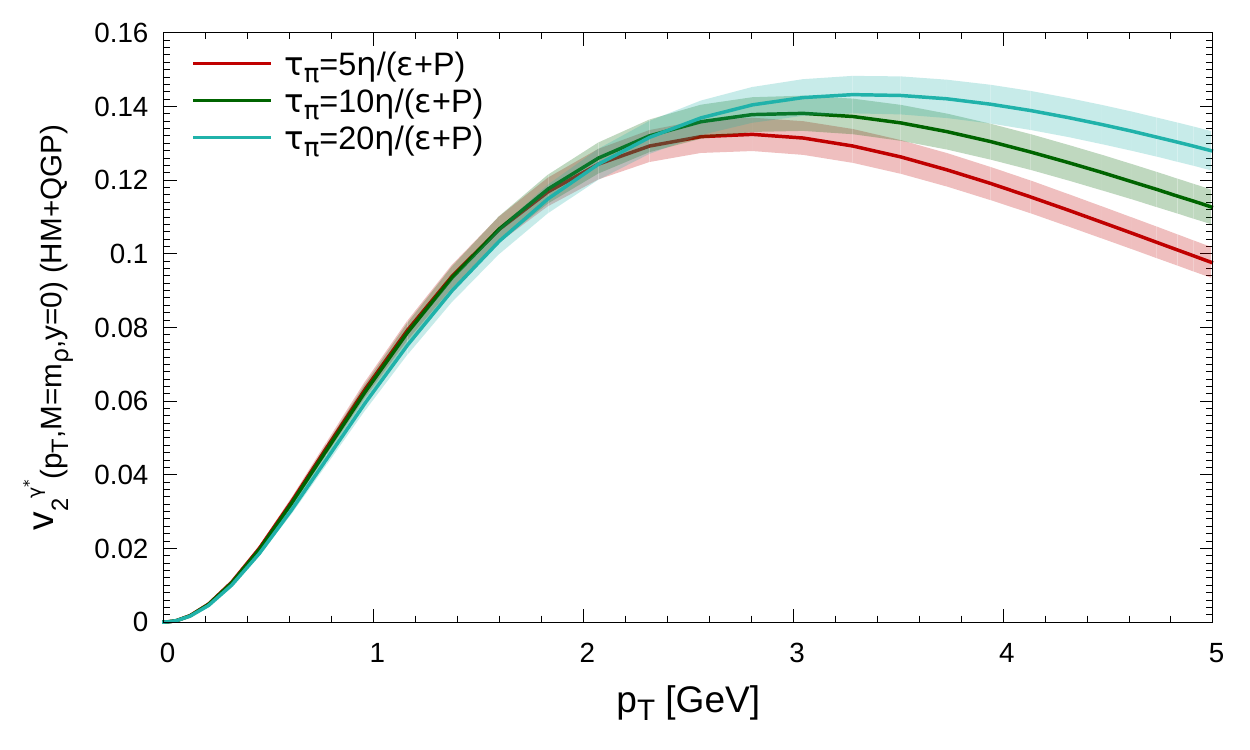}}
\subfigure[]{\includegraphics[width=0.497%
\textwidth]{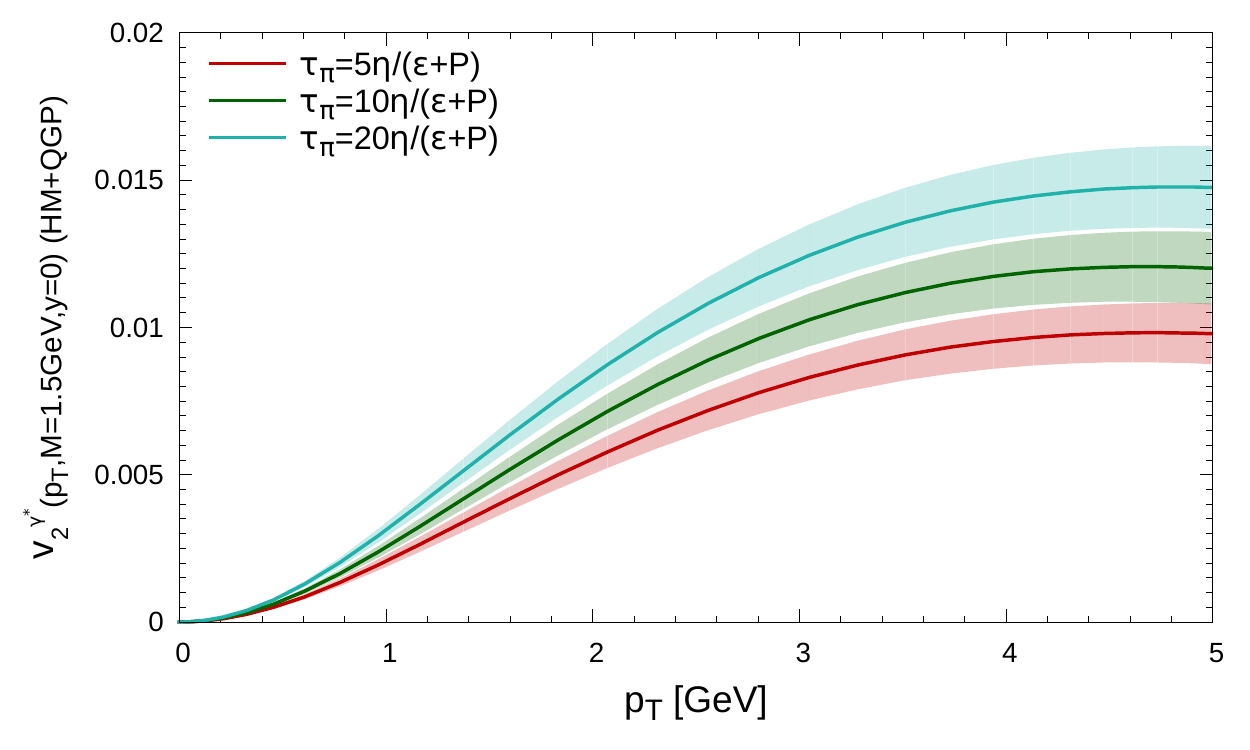}}
\caption{(Color online) Differential elliptic flow of thermal dileptons for
(a) $M=m_\protect\rho$ and (b) $M=1.5$GeV as a function of transverse
momentum, for different values of shear relaxation time.}
\label{fig:thermal_dilepton_v2_pt_M}
\end{figure}

The differential elliptic flow of thermal dileptons $v_{2}(p_{T})$ [Fig. \ref%
{fig:thermal_dilepton_v2_pt_M} (a)], for a low invariant mass $M=m_\rho$ ---
where HM dileptons dominate --- is affected by $\tau_\pi$ in manner similar
to thermal photon's $v_2(p_T)$ [Fig. \ref{fig:thermal_photon} (b)]. The flow
of intermediate mass dileptons in Fig.~\ref{fig:thermal_dilepton_v2_pt_M}
(b), where QGP emission is the main source, has an increased sensitivity to $%
\tau_\pi$; consistent with Fig.~\ref{fig:photon_hg_qgp_v2} (a). The effect
of $\tau_\pi$ is not limited to $v_2(p_T)$ and is also affecting higher flow
harmonics in a similar fashion, as can be seen Fig.~\ref%
{fig:dilepton_v3_v4_pt}.

\begin{figure}[!h]
\begin{center}
\subfigure[]{\includegraphics[width=0.497%
\textwidth]{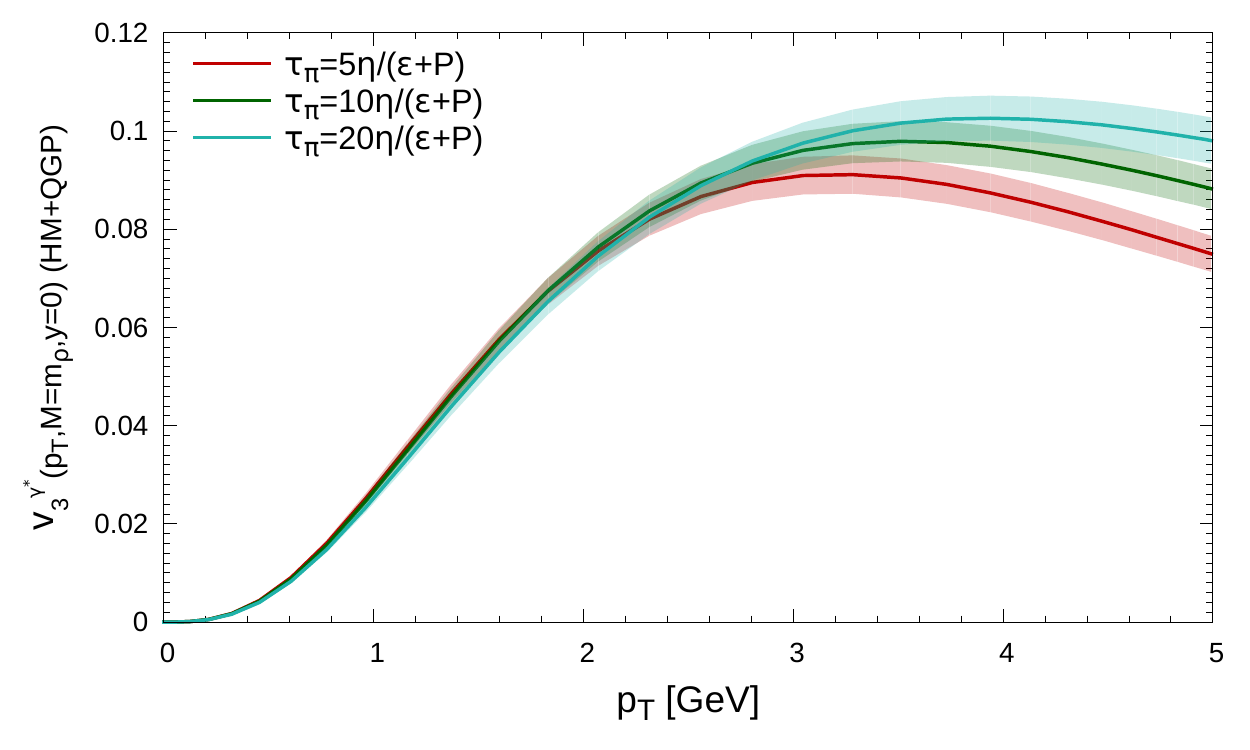}}
\subfigure[]{\includegraphics[width=0.497%
\textwidth]{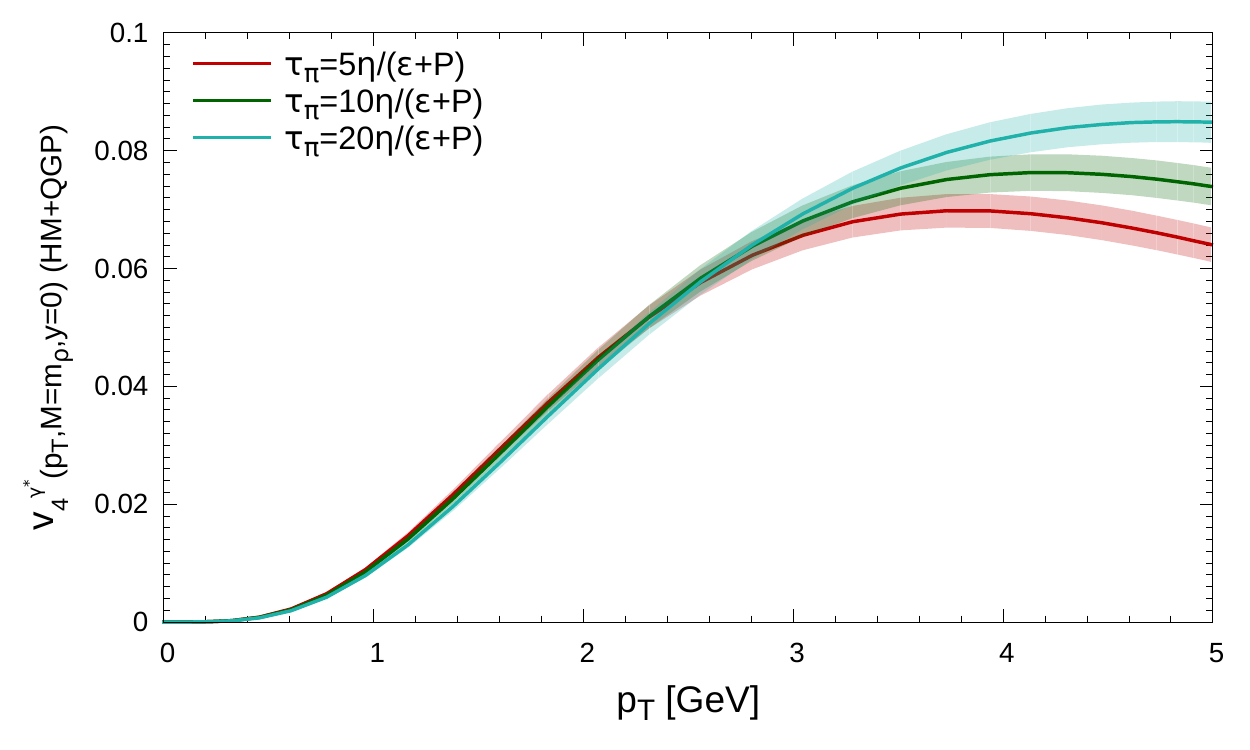}}
\end{center}
\caption{(Color online) Influence of $\protect\tau_\protect\pi$ on higher
flow harmonics of thermal dileptons at $M=m_\protect\rho$: (a) $v_3(p_T)$
(b) $v_4(p_T)$.}
\label{fig:dilepton_v3_v4_pt}
\end{figure}

\begin{figure}[!h]
\centering
\subfigure[]{\includegraphics[width=0.497%
\textwidth]{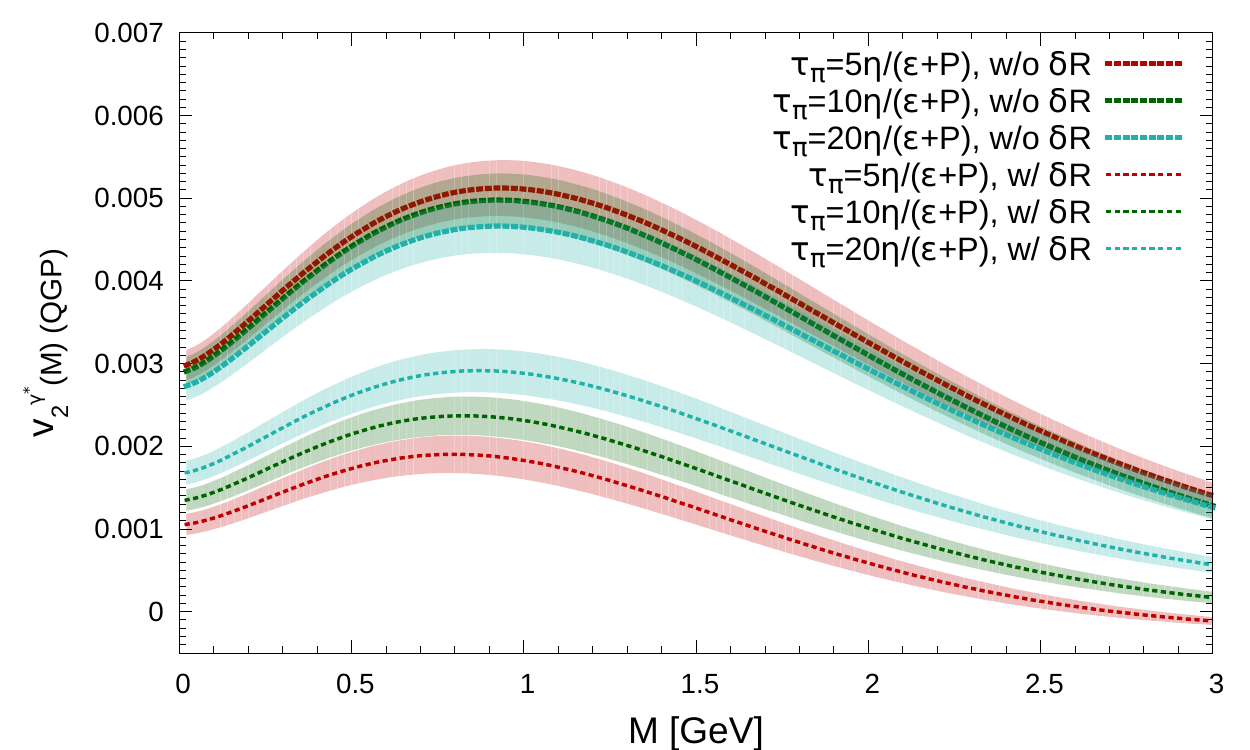}} %
\subfigure[]{\includegraphics[width=0.497%
\textwidth]{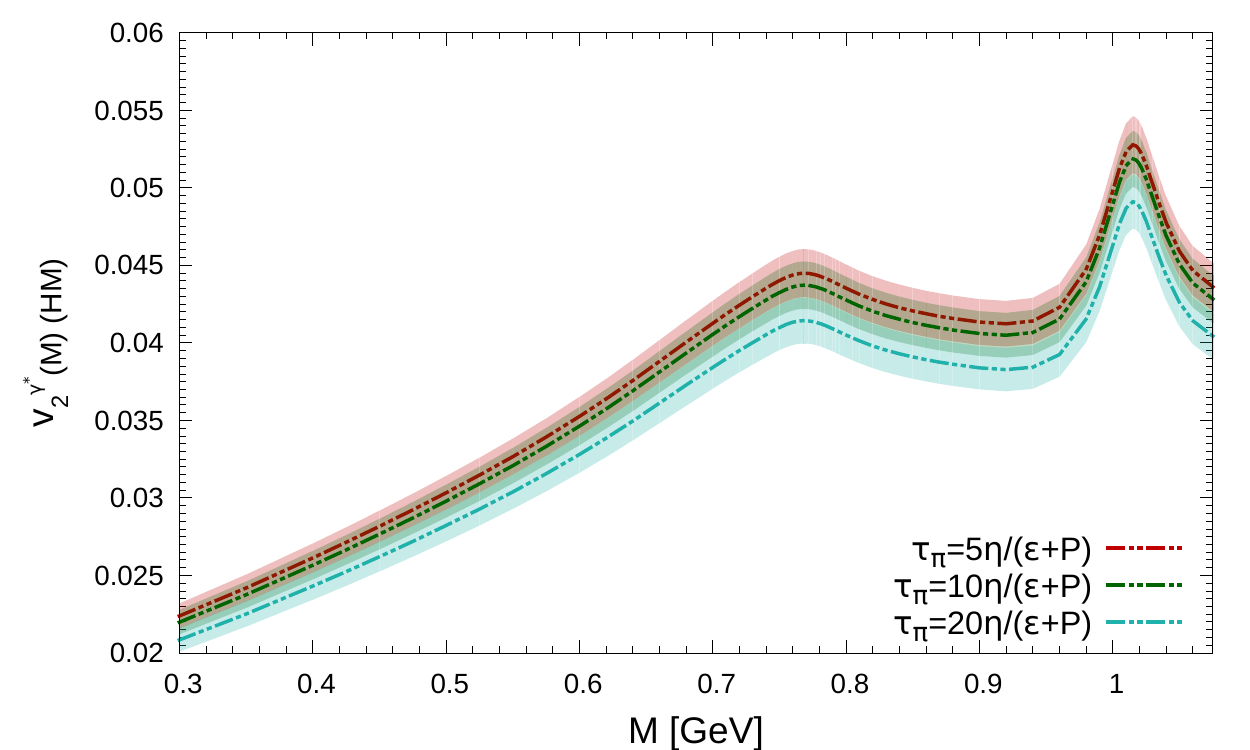}}
\caption{(Color online) Differential elliptic flow of dileptons emitted by
the QGP (a) and emitted by the hadronic medium (b) as a function of the
dilepton invariant mass, for different values of shear relaxation time. The
effects of the viscous corrections to the QGP rate are presented in (a)
whereas those of the HM rate are small and hence only results using the rate
including viscous corrections is presented.}
\label{fig:dilepton_hg_qgp_v2_M}
\end{figure}

In Fig.~\ref{fig:dilepton_hg_qgp_v2_M} (a) and (b) we show the elliptic flow
of dileptons emitted from the QGP and HM, respectively, as a function of the
dilepton invariant mass. We see that the elliptic flow of dileptons emitted from
the QGP increases with $\tau _{\pi }$ while the opposite behavior is seen
for dileptons from the HM. This effect can be better understood by first
analyzing the elliptic flow without any viscous corrections to the rates, $%
\delta R$, shown in Fig.~\ref{fig:dilepton_hg_qgp_v2_M}\footnote{Though the effect of $%
\delta R$ is negligible on the $v_2(M)$ of HM dileptons, it is included in Fig.~\ref{fig:dilepton_hg_qgp_v2_M} (b).}.
In the case without $\delta R$ corrections, we see that the $v_{2}$ of dileptons from both the QGP and HM
actually decreases with increasing relaxation time. This happens because, at
early times, the shear stress-tensor increases the transverse pressure,  
leading to a larger anisotropic flow. Since systems with smaller relaxation time 
develop larger values of shear stress tensor at earlier times (see Fig. %
\ref{fig:pi_munu_tau_pi}), they will also have larger values of elliptic
flow. On the other hand, the viscous corrections to the rate reduces the
elliptic flow and is proportional to the shear-stress tensor. Therefore,
smaller relaxation times will generate a larger reduction of elliptic
flow due to $\delta R$. This effect is very large in the QGP phase and ends
up reverting the previously observed trend.

\begin{figure}[h]
\centering
\subfigure[]{\includegraphics[width=0.497%
\textwidth]{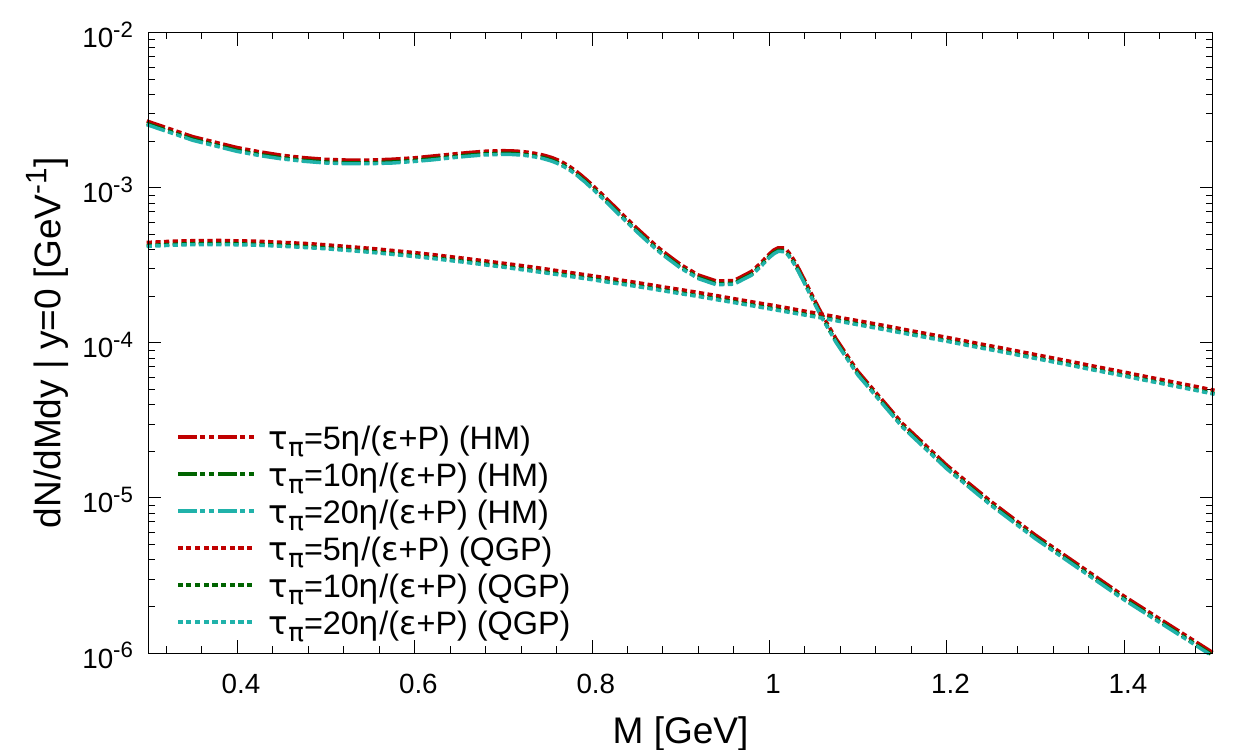}}
\subfigure[]{\includegraphics[width=0.497%
\textwidth]{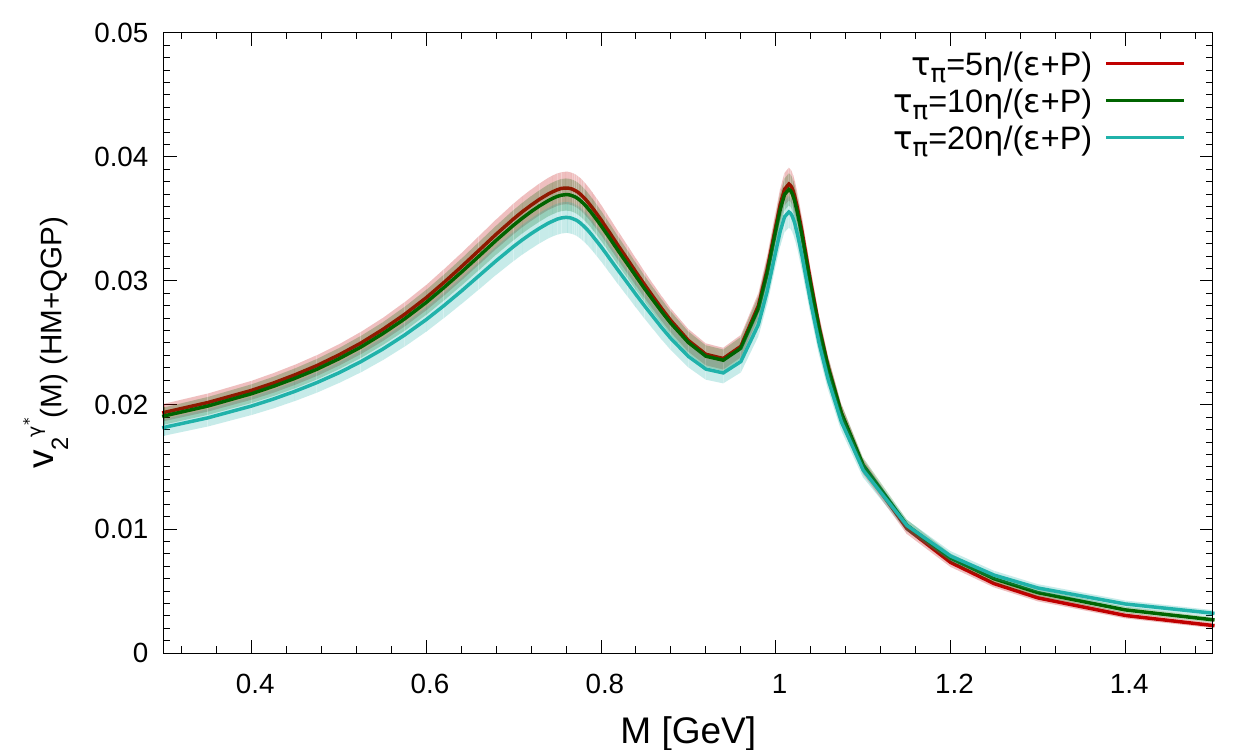}}
\caption{(Color online) Dilepton yield (a) and elliptic flow (b) as a
function of invariant mass, for different values of shear relaxation time.}
\label{fig:dileptons_hg_qgp_yield_thermal_v2_M}
\end{figure}

Lastly, in Fig.~\ref{fig:dileptons_hg_qgp_yield_thermal_v2_M} we show the
total (QGP+HM) thermal dilepton elliptic flow as a function of invariant
mass, for the three different shear relaxation times. At small invariant
masses, $M<1.1$ GeV, the HM dileptons are dominant and we see that the
elliptic flow is reduced as the relaxation time increases. On the other
hand, for larger invariant masses, $M>1.1$ GeV, the QGP contribution starts
to dominate and the dependence on the relaxation time is inverted. This
behavior is expected and is in agreement with the one observed for thermal
photons. Note that the invariant mass over which this behavior switches
(here, $M\approx 1.1$ GeV) is not universal and depends on other parameters,
such as the freeze-out temperature and the initialization time. If one
starts the simulation earlier, more QGP thermal photons/dileptons can be
emitted, while if one decreases the freeze-out temperature, more hadron gas
photons/dileptons are emitted. In fact, because of the initial and
freeze-out conditions chosen in this study, the net effect of $\tau_\pi$ on
the total thermal $v_2(M)$ is not large as there are incomplete
cancellations between the behavior in the QGP and HM sectors. So, one should
always take into account the initial and freeze-out conditions when
interpreting results of thermal EM emissions in heavy-ion collision
simulations.


\subsection{Effect of an initial shear-stress tensor}

\label{discussionInit}

We explore in this section the sensitivity of EM probes to the presence of a
non-vanishing initial shear-stress tensor. As already stated, we use a
rescaled Navier-Stokes value of $\pi ^{\mu \nu }$ as initial condition:

\begin{equation*}
\pi ^{\mu \nu }(\tau _{0})=c\times \mathrm{diag}\left( 0,\frac{2\eta }{3\tau
_{0}},\frac{2\eta }{3\tau _{0}},-\frac{4\eta }{3\tau _{0}}\right) .
\end{equation*}%
We use three different values of $c$=0, 1/2, 1, with the case $c=1$
corresponding to the Navier-Stokes limit, $c=0$ to the equilibrium limit,
and $c=1/2$ to an intermediary case. We set $\tau _{\pi }=5\eta
/(\varepsilon +P)$ for this whole section, and generate 200 hydrodynamical events for each value of $c$. As we saw in the previous
section, the choice of $\tau _{\pi }$ is important since it determines the
timescale for $\bar{\pi}^{\mu \nu }=\pi ^{\mu \nu }/(\varepsilon +P)$ to
converge to the Navier-Stokes limit from the chosen initial conditions. 
\begin{figure}[h]
\begin{center}
\includegraphics[width=0.497\textwidth]{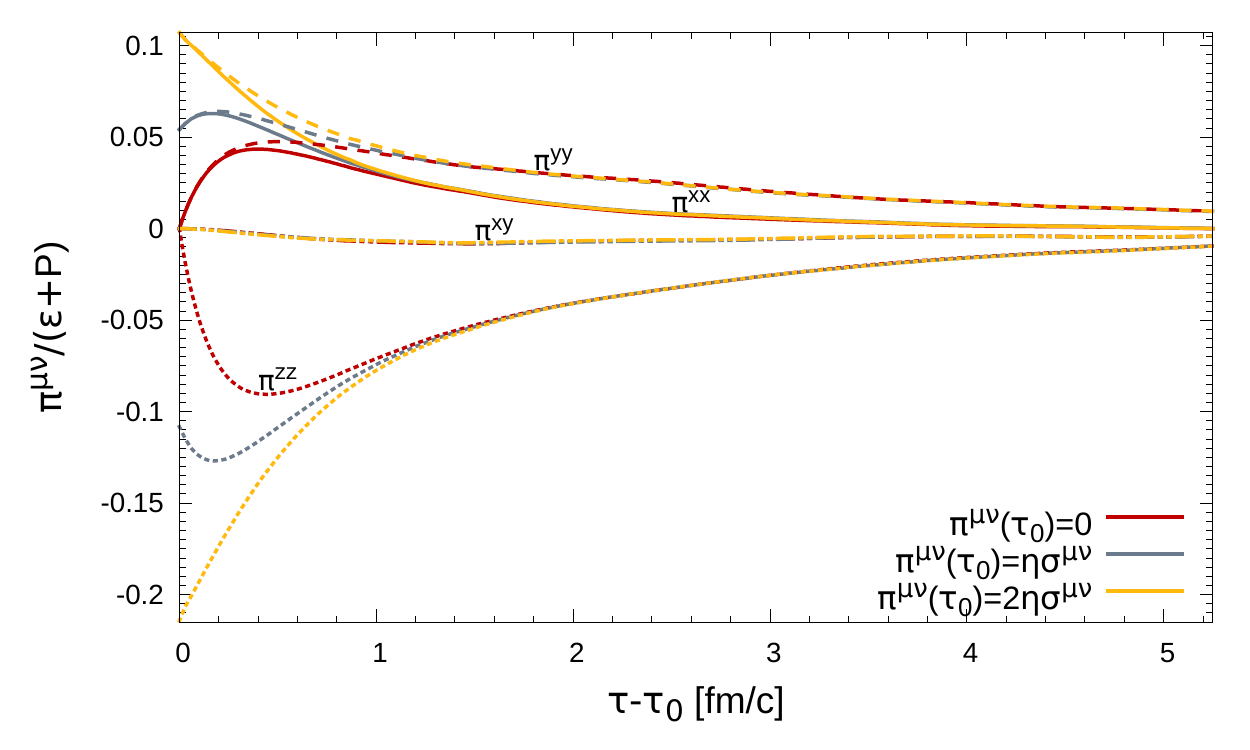}
\end{center}
\caption{(Color online) Shear-stress tensor for $c=0,1/2,1$ in the local
rest frame of the fluid cell located at x=y=2.625 fm, z=0 fm, averaged over
all events. Results with $c=0$ are displayed in red, $c=1/2$ in gray, and $%
c=1$ in yellow.}
\label{fig:init_pi_munu}
\end{figure}

\begin{figure}[!h]
\subfigure[]{\includegraphics[width=0.497%
\textwidth]{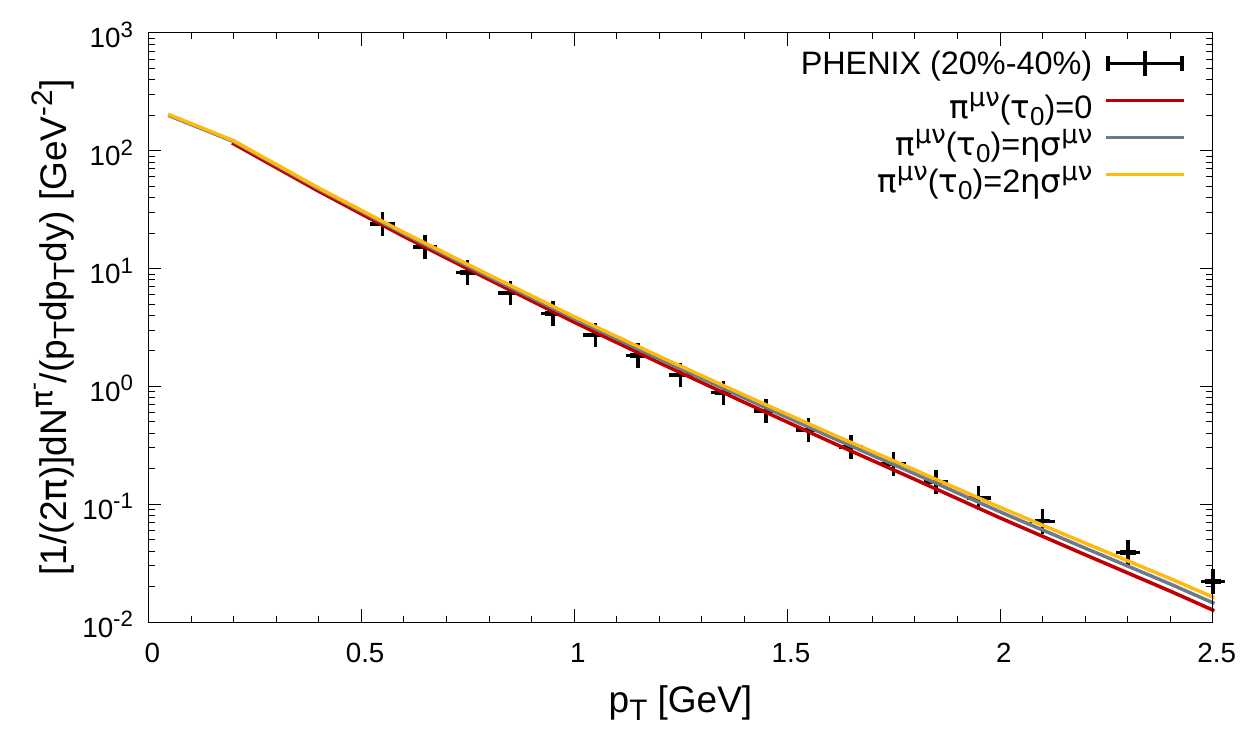}} %
\subfigure[]{\includegraphics[width=0.497%
\textwidth]{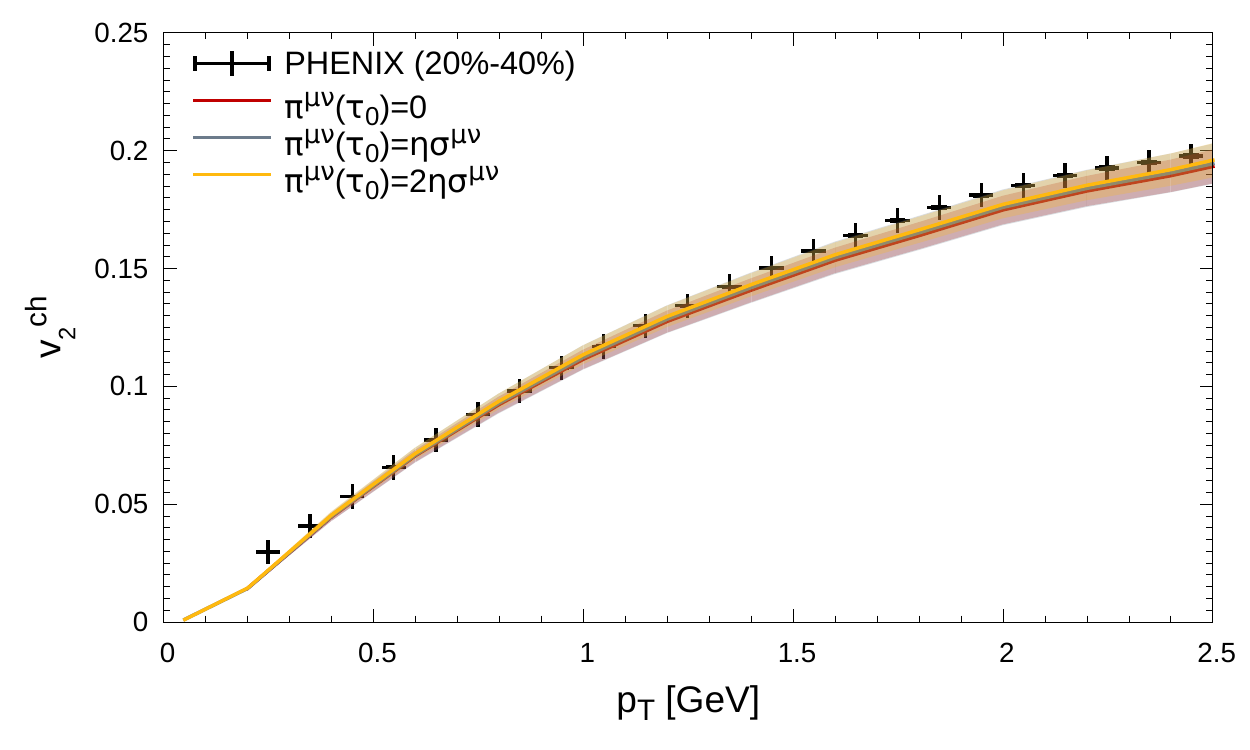}}
\caption{(Color online) Pion transverse momentum spectra (a) and charged
hadron differential elliptic flow (b) as a function of transverse momentum,
for different values of the initial shear-stress tensor.}
\label{fig:hadrons2}
\end{figure}

We show in Fig.~\ref{fig:init_pi_munu} the time dependence of various
components of $\bar{\pi}^{\mu \nu }(\tau )$, in the rest frame of the fluid,
for our three different choices of initial conditions. Notice that
differences in $\bar{\pi}^{\mu \nu }$ at early times are washed out within $%
\sim $1.5 fm/c, which is about a quarter of the medium's lifetime depicted
in Fig.~\ref{fig:init_pi_munu}. This implies that hadrons should be largely
insensitive to changes in the initial $\bar{\pi}^{\mu \nu }$, though photons
and dileptons produced early enough in the collision could be sensitive to
the different initial conditions. The spectra and $v_{2}$ of hadrons (Fig. %
\ref{fig:hadrons2}) agrees with our interpretation of Fig.~\ref%
{fig:init_pi_munu}, with both observables showing a very weak dependence on
the initial $\pi ^{\mu \nu }$.

\begin{figure}[h]
\subfigure[]{\includegraphics[width=0.497%
\textwidth]{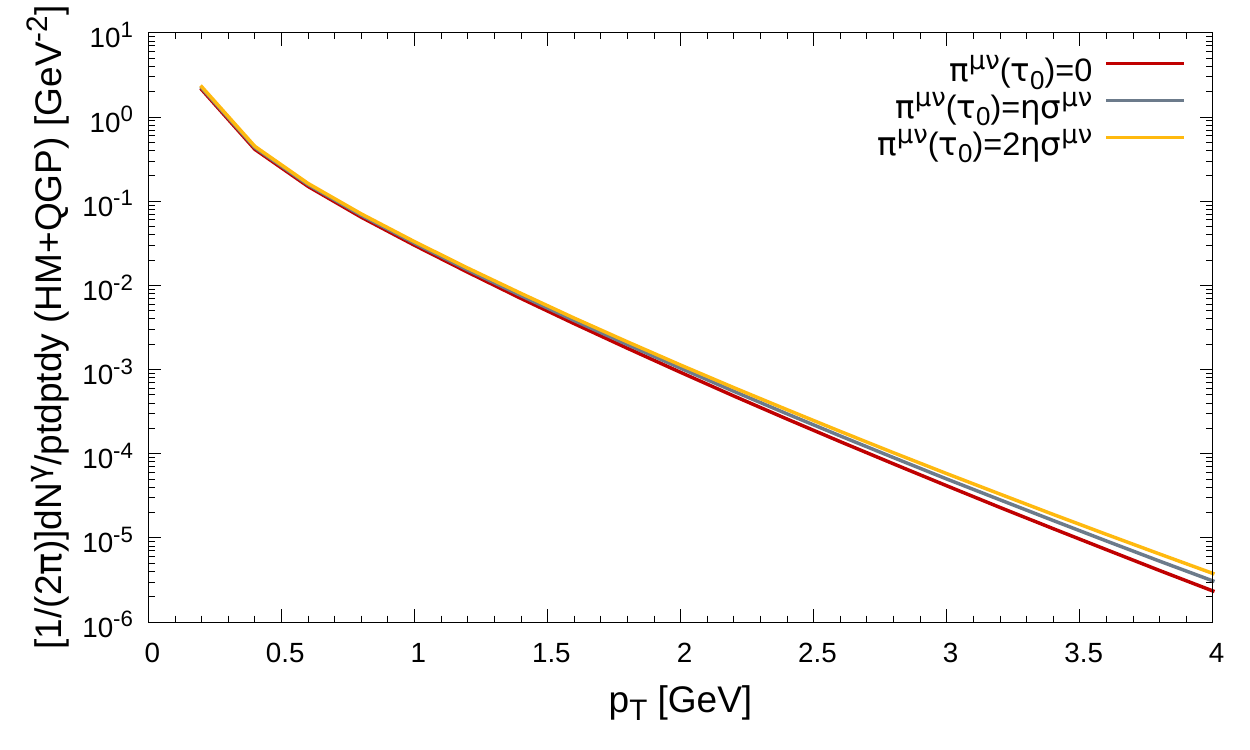}}
\subfigure[]{\includegraphics[width=0.497%
\textwidth]{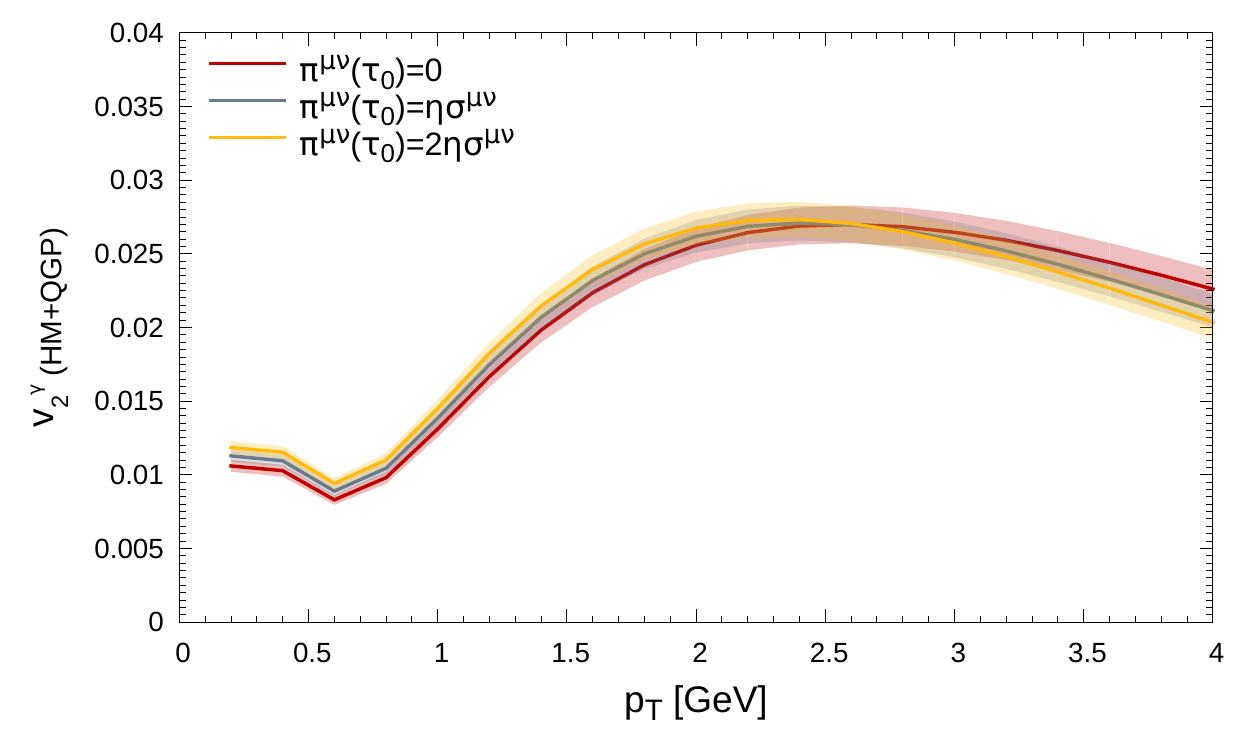}}
\caption{(Color online) Transverse momentum spectra (a) and differential
elliptic flow (b) of thermal photons as a function of transverse momentum,
for different values of initial $\protect\pi ^{\protect\mu \protect\nu }$.}
\label{fig:photon4}
\end{figure}
The photon spectra and $v_{2}$ are shown on Fig. \ref{fig:photon4} (a) and
(b), respectively. Similar to what was seen on hadrons, the effect of the
initial $\pi ^{\mu \nu }$ on the photon spectra is small. However, initial $%
\pi ^{\mu \nu }$ does change the shape of photon $v_{2}$; a small increase
at low $p_{T}$ is observed, while the reverse behavior occurs at high $p_{T}$%
. 
\begin{figure}[h]
\centering 
\includegraphics[width=0.497%
\textwidth]{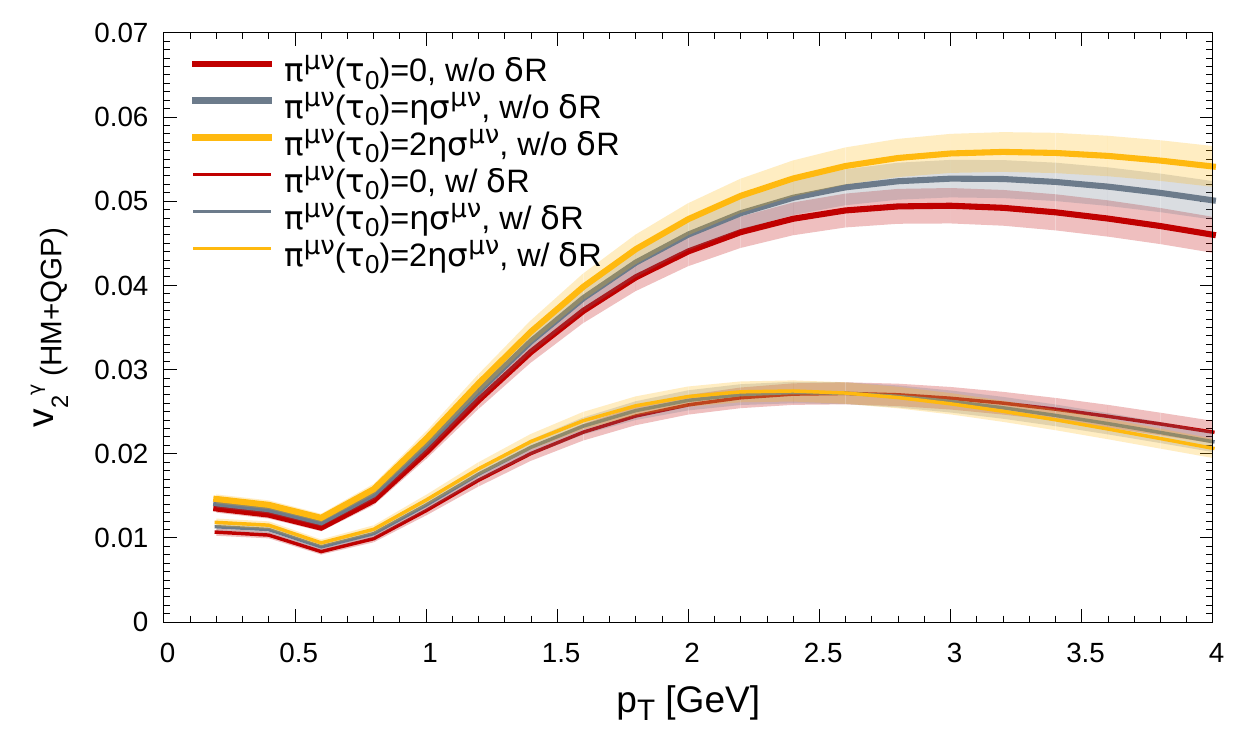}
\caption{(Color online) Thermal photon elliptic flow with and without
viscous corrections ($\protect\delta $R) to the emission rates.}
\label{fig:photon5}
\end{figure}

To better understand the origin of these features, we show in Fig.~\ref%
{fig:photon5} the $v_2$ of photons with and without the effect of $\delta$R
to the photon production rate. Recall that the former will only be sensitive
to the feedback of initial $\pi^{\mu\nu}$ on the temperature and flow
profiles, while the latter contains the direct effect of the change in $%
\pi^{\mu\nu}$ on the photon production rate. Indeed, the effect of the
initial $\pi^{\mu\nu}$ on the hydrodynamical evolution is not small, and
alone produces a significant increase on the photon $v_2$ (see Fig. \ref%
{fig:photon5}). The change of behavior seen at high $p_T$ is thus solely due
to the effect of the viscous $\delta R$ correction to the photon production
rate. It is more apparent at high $p_T$ because the viscous correction to
the rate is larger in that region, relative to low $p_T$.

The $p_T$ dependence of dilepton's elliptic flow at $M=m_\rho$ is similar to
photon's $v_2(p_T)$, as was noted in section~\ref{discussionTau}, while the $%
v_2(p_T)$ at higher invariant masses is small. Furthermore, higher flow
harmonics of thermal dileptons as a function of $p_T$ at $M=m_\rho$ are also
affected in very similar way to $v_2$. The same statement holds true for
higher flow harmonics of thermal photons. Hence, we focus immediately on the
thermal dilepton invariant mass distribution.

\begin{figure}[!h]
\centering
\subfigure[]{\includegraphics[width=0.497%
\textwidth]{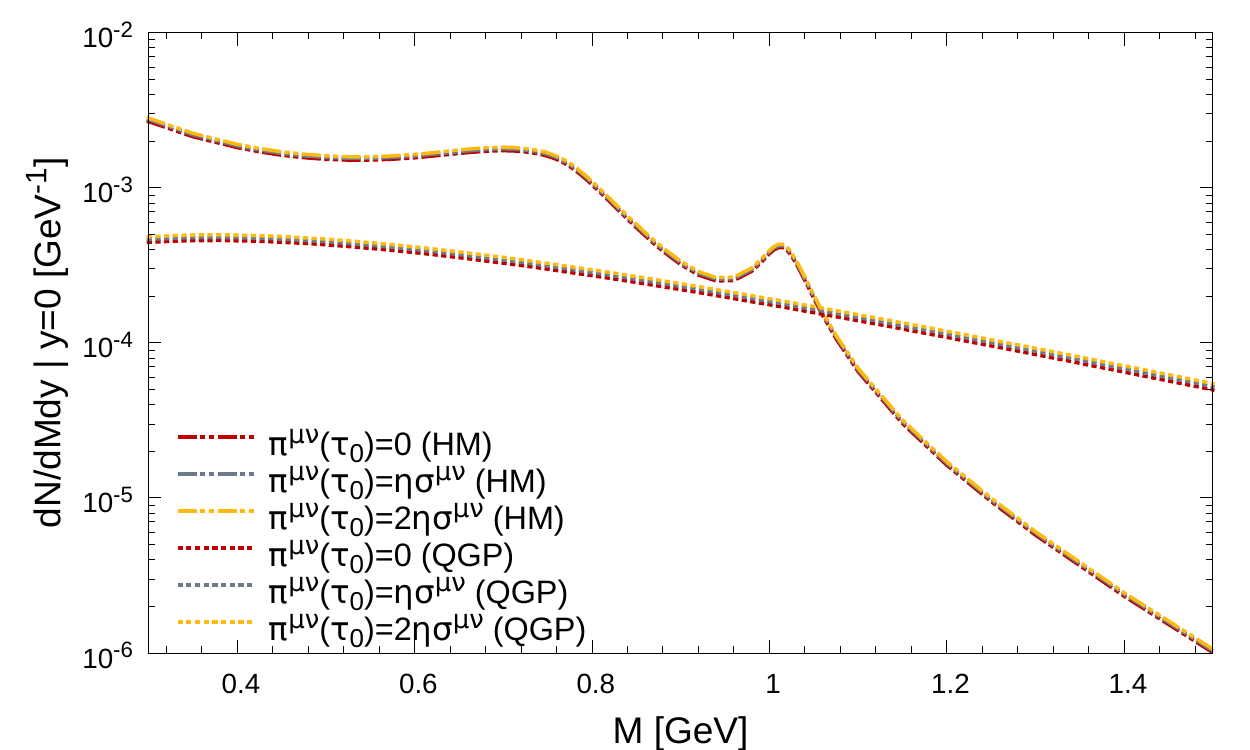}}
\subfigure[]{\includegraphics[width=0.497%
\textwidth]{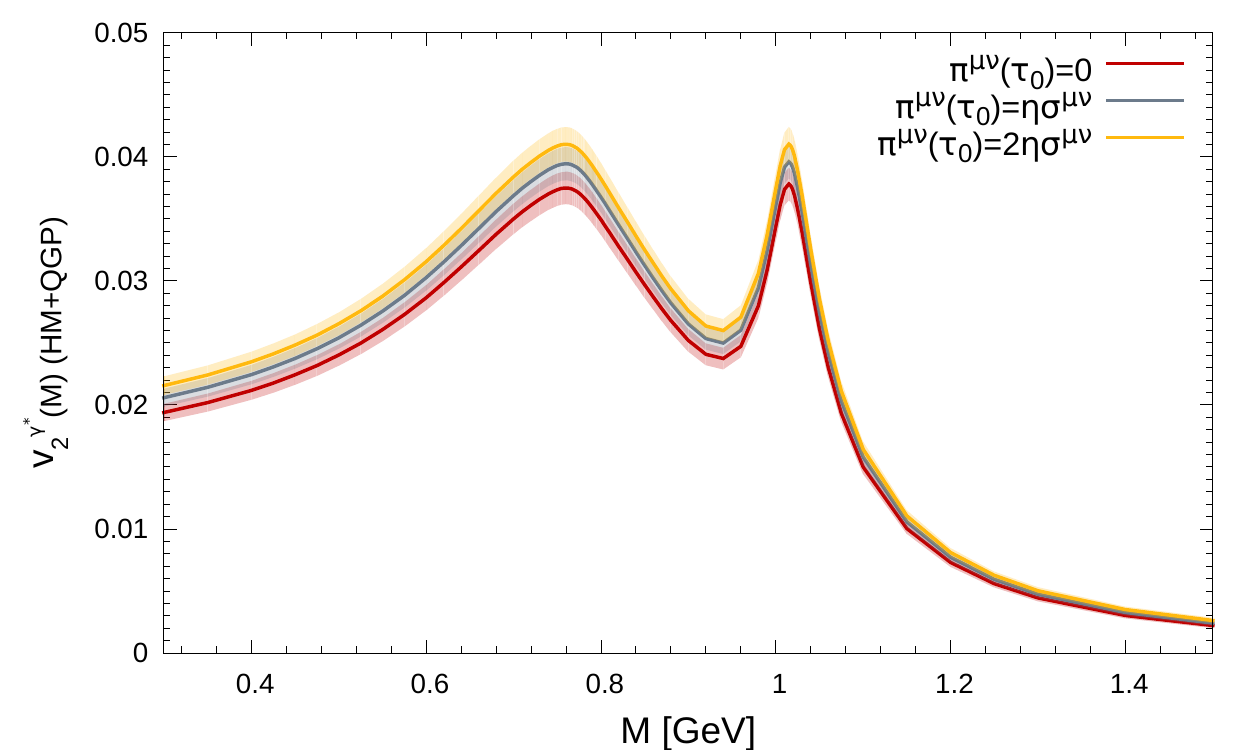}}
\caption{(Color online) Dilepton yield (a) and elliptic flow (b) as a
function of invariant mass, for different values of initial $\protect\pi^{%
\protect\mu\protect\nu}$.}
\label{fig:dileptons4}
\end{figure}

The invariant mass yield of thermal dileptons doesn't depend on any viscous
corrections\footnote{%
After performing a tensor decomposition on $\delta R$, and integrating over
the 3-momentum $\mathbf{q}$, the only tensor that can be constructed is
proportional to $u_\mu u_\nu$ which vanishes when contracted with $%
\pi^{\mu\nu}$. Hence the invariant mass distribution of dilepton yield must
be independent of $\delta R$.} and hence it is only sensitive to the entropy
generation that a non-zero $\pi^{\mu\nu}(\tau_0)$ injects into the system,
which is small as can be seen in Fig. \ref{fig:dileptons4} (a). Also, since
the invariant mass yield is unaffected by $\delta R$, $v_2(M)$ from both
thermal sources behaves in a more monotonic fashion as $\pi^{\mu\nu}(\tau_0)$
increases [see Fig. \ref{fig:dileptons4} (b)].

\begin{figure}[!h]
\centering
\subfigure[]{\includegraphics[width=0.497%
\textwidth]{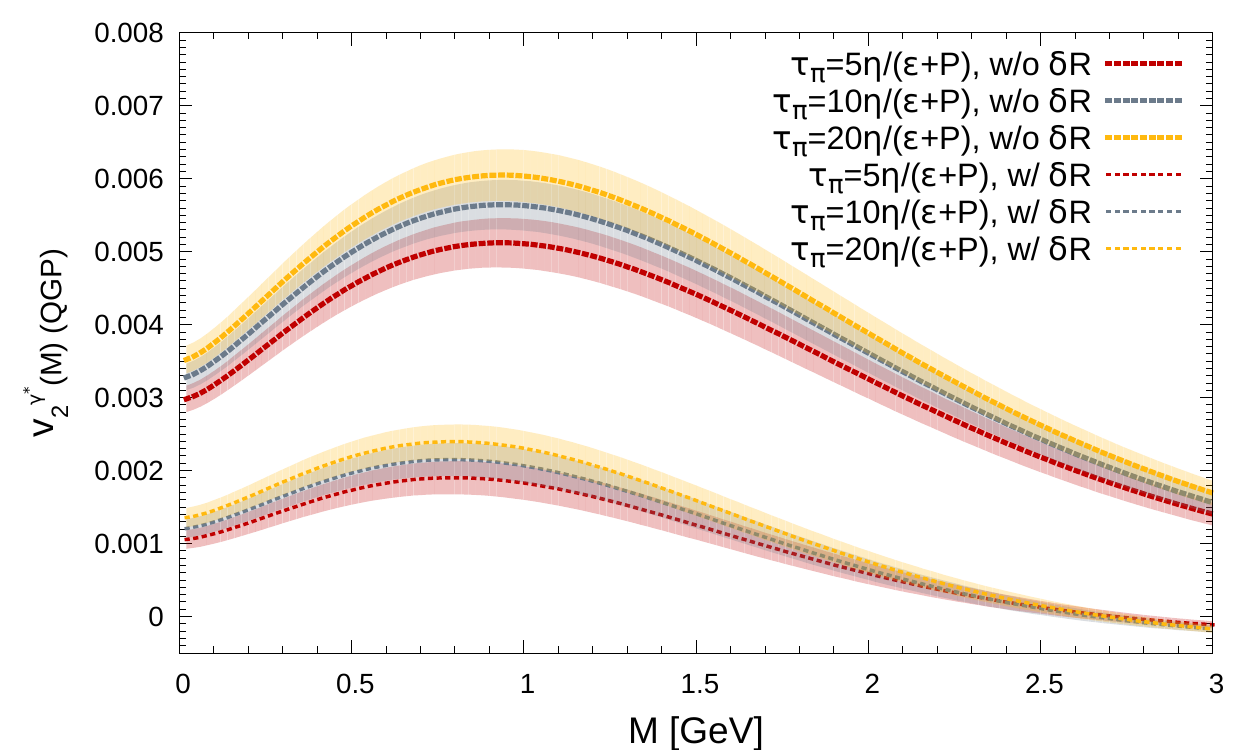}} %
\subfigure[]{\includegraphics[width=0.497%
\textwidth]{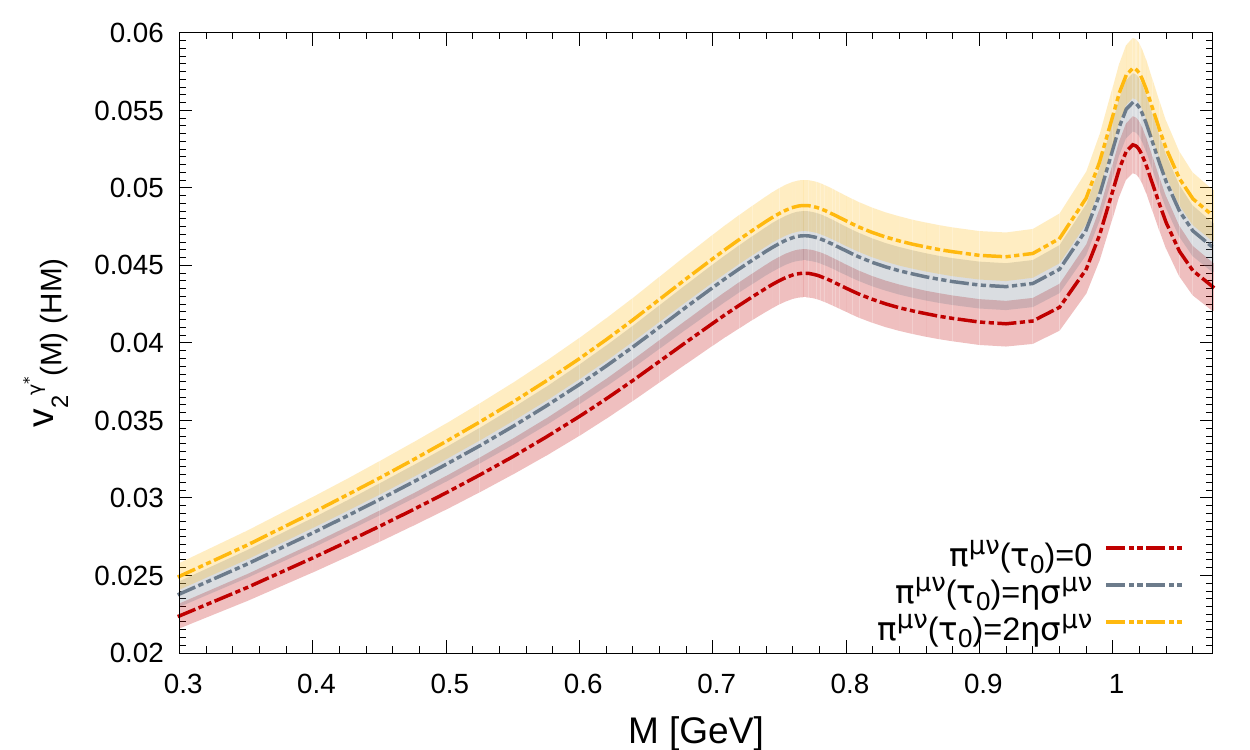}}
\caption{(Color online) Differential elliptic flow of dileptons emitted by
the QGP (a) and emitted by the hadronic medium (HM) (b) as a function of the
dilepton invariant mass, for different values of initial shear stress
tensor. Only the QGP dileptons are calculated with and without viscous
corrections $\protect\delta$R to the rate, while the HM dileptons are
calculated with viscous $\protect\delta R$ corrections.}
\label{fig:dileptons5}
\end{figure}

Similarly to photon's $v_{2}(p_{T})$ without $\delta $R, $v_{2}(M)$ for the
QGP in Fig. \ref{fig:dileptons5} (a) increases with $\pi ^{\mu \nu }(\tau
_{0})$, while the viscous corrections are mostly reducing the $v_{2}$. The
shape of the $v_{2}(M)$ changes somewhat at higher $M$ as $\pi ^{\mu \nu
}(\tau _{0})$ increases owing to $\delta R$ effects in the numerator of $%
v_{2}$, however those viscous corrections are not inverting the order of the 
$v_{2}(M)$ curves, as was the case for photons. The dilepton HM sector
behaves monotonically as a function of M as $\pi ^{\mu \nu }(\tau _{0})$
increases as shown in Fig. \ref{fig:dileptons5} (b), receiving an increase
of at most $\sim 10$\% by the time $\pi ^{\mu \nu }(\tau _{0})=2\eta \sigma
^{\mu \nu }$. Hence, $v_{2}(M)$ of QGP and HM are directly exposing the
modifications of the hydrodynamical evolution, which is seen as a definite 
trend as far as their sensitivity to $\pi ^{\mu \nu }(\tau _{0})$ is concerned. 
In fact, this trend is preserved for dileptons when going to even higher 
$\pi^{\mu \nu }(\tau _{0})$, namely $\pi ^{\mu \nu }(\tau _{0})=4\eta \sigma
^{\mu \nu }$, while hadrons remain unaffected, as was shown in Ref. 
\cite{Vujanovic:2014xva}, where a simpler optical Glauber initial condition 
was used. %

Going back to the Fig. \ref{fig:init_pi_munu}, the large pressure gradients
in the longitudinal direction are more significantly reduced via a non-zero
initial $\bar{\pi}^{zz}(\tau_0)$ --- relative to $\bar{\pi}^{zz}(\tau_0)=0$ --- and 
thus are more efficiently transferred onto the transverse plane. This coupling 
of the longitudinal and transverse pressure gradients causes an increase in 
the $v_2(M)$ of QGP dileptons. The elliptic flow $v_2(M)$ of HM dileptons is 
also increased owing to the fact that a cross-over phase transition allows for 
HM dileptons to be emitted from early times before $\bar{\pi}^{\mu\nu}$ has 
relaxed to the Navier-Stokes value.
%

\section{Conclusions}

\label{Sec:Conclusion} 

In this paper we studied the effect of the shear relaxation time on thermal
photons and dileptons emitted from the QCD medium created at the top RHIC
energy. We further analyzed how initial conditions, more specifically
initial $\pi^{\mu\nu}(\tau_0)$, of the fluid-dynamical description affect
thermal EM probes. We concluded that thermal photons and dileptons can be
sensitive to $\tau_\pi$ and to initial conditions of $\pi^{\mu\nu}$ used in
the modeling of the collision, while hadronic observables are poorly sensitive 
to those two parameters.%

We have shown that the shear relaxation time has a visible effect on thermal
photon and dilepton elliptic flow, with larger values of relaxation time
leading to an increase in photon and dilepton $v_2(p_T)$. This indicates
that thermal EM probes could be used in the future to provide constraints on
the size of the relaxation time for $\pi^{\mu\nu}$ in QCD matter. We further
computed higher flow harmonics of thermal dileptons, and have shown that the
same effects $\tau_{\pi}$ induces on $v_2$ also persist for $v_3$ and $v_4$.
In addition, we demonstrated that thermal EM radiation is sensitive to the
initial conditions of hydrodynamics, specifically the initial shear-stress
tensor $\pi^{\mu\nu}(\tau_0)$. In particular, the elliptic flow of thermal
dileptons as a function of invariant mass has a definite trend: it increases
the elliptic flow with larger $\pi^{\mu\nu}(\tau_0)$.

While larger values of relaxation time and $\pi^{\mu\nu}(\tau_0)$ affect the
elliptic flow of thermal photons, the effect is mild except at high
transverse momentum, where prompt photons dominate over thermal ones. There
is also barely any effect on the thermal photon spectra. In consequence, it
does not appear that the effects investigated in this work would
significantly change the agreement with direct photon data of current
hydrodynamical calculations (e.g. Ref.~\cite{Paquet:2015lta}). It would
nevertheless be interesting to revisit these effects as more realistic
initial conditions and more complete viscous corrections to the thermal
emission rates become available. 

Similarly, the contribution of open heavy flavor hadron pairs, whose
semi-leptonic decay contribute to dileptons, needs to be included to the
list of dilepton sources. Indeed, it was recently shown that open charm
hadrons \cite{Vujanovic:2013jpa} traveling through the medium develop flow
which contributes to dilepton $v_2(M)$ in the intermediate mass region\footnote{Drell-Yan processes and the decay of heavy quarkonium also 
contribute to the total dilepton spectrum. However, in the low and intermediate 
mass regions, their contribution is usually small enough to be neglected.}. 
In the low mass region, late decays of pseudo-scalar 
mesons and $\omega$ and $\phi$ mesons all contribute to the dilepton 
``cocktail'' production. However, in the invariant mass window $0.6<M<0.78$ 
GeV, thermal dilepton production is the dominant source of lepton pairs \cite%
{Vujanovic:2013jpa}, and thus results presented here should persist even
when the ``cocktail'' and open heavy flavor contributions are added. A more
in-depth study including the ``cocktail'' and open heavy flavor dileptons is
in progress.

After taking into account all these sources, extracting transport
coefficients from EM probes will remain a challenging task, but we are
optimistic it can be led to a fruitful completion. In the case of dileptons,
the task might be made easier if heavy flavor tracking is used to remove the
open heavy flavor signal from the measured dilepton flow in the intermediate
mass and low mass regions. Performing this subtraction in experimental
dilepton data opens an interesting ``window'' to extract the transport
coefficients of QGP in the intermediate mass region. In the low mass region,
removing the dilepton ``cocktail'' will help study transport coefficients
coming from a low temperature QCD medium. On the photon side, once good
agreement with direct photon measurements is achieved, one interesting
avenue to isolate the thermal contribution may lie in taking ratios of
anisotropic coefficients, making thermal photons stand out from sources that
have small/negligible flow anisotropies \cite{Shen:2014cga}. These would all
strengthen the capabilities of photons and dileptons as complementary probes
of the properties of QCD media, especially of non-equilibrium properties of the initial
conditions. 


\section*{Acknowledgments}

G.V. is grateful to B. Schenke for helpful discussions. This work was
supported in part by the Natural Sciences and Engineering Research Council
of Canada, and in part by the Director, Office of Energy Research, 
Office of High Energy and Nuclear Physics, Division of Nuclear Physics, of 
the U.S. Department of Energy under Contracts No. DE-AC02-98CH10886, DE-AC02-05CH11231, 
and \rm{DE-SC0004286}. G. Vujanovic acknowledges support by the Fonds Qu\'eb\'ecois 
de Recherche sur la Nature et les Technologies (FQRNT), the Canadian Institute 
for Nuclear Physics, and by the Seymour Schulich Scholarship, while G.S. Denicol 
acknowledges support through a Banting Fellowship from the Government of Canada. 
Computations were performed on the Guillimin supercomputer at McGill University 
under the auspices of Calcul Qu\'ebec and Compute Canada. The operation of 
Guillimin is funded by the Canada Foundation for Innovation (CFI), the National 
Science and Engineering Research Council (NSERC), NanoQu\'ebec, and the Fonds 
Qu\'eb\'ecois de Recherche sur la Nature et les Technologies (FQRNT).

\bibliography{references}


\end{document}